%% file: paper_final.tex
\documentclass[fleqn,usenatbib]{mnras}

\usepackage{newtxtext,newtxmath}
\usepackage[T1]{fontenc}
\usepackage{ae,aecompl}

\usepackage{graphicx}	% Including figure files
\usepackage{longtable}
\usepackage{booktabs} 

\newcommand{\kepler}{{\em Kepler\/}}
\newcommand{\tess}{{\em TESS\/}}
\newcommand{\dsct}{\mbox{$\delta$~Sct}}
\newcommand{\gdor}{\mbox{$\gamma$~Dor}}
\newcommand{\cd}{\mbox{d$^{-1}$}}

\DeclareMathOperator{\sinc}{sinc}
\newcommand{\echelle}{{\'e}chelle}
\newcommand{\Dnu}{\mbox{$\Delta\nu$}}

\newcommand{\new}[1]{{\color{red}\textbf{#1}}}
\renewcommand{\new}[1]{#1} % for arxiv version

\title[Revisiting bright $\delta$ Scuti stars]{Revisiting bright $\delta$ Scuti stars and their period--luminosity relation with \tess\ and {\em Gaia} DR3}

\author[Barac et al.]{%
Natascha Barac$^{1,2}$\thanks{E-mail: nbarac@sas.upenn.edu}, 
Timothy R. Bedding$^{2,3}$\thanks{E-mail: tim.bedding@sydney.edu.au},
Simon J. Murphy$^{2,3,4}$ and
Daniel R. Hey$^{2,3,5}$
\\
$^1$Department of Physics and Astronomy, University of Pennsylvania, Philadelphia, PA 19104, USA \\
$^2$Sydney Institute for Astronomy, School of Physics, University of Sydney, NSW 2006, Sydney, Australia \\
$^3$Stellar Astrophysics Centre, Department of Physics and Astronomy, Aarhus University, DK-8000 Aarhus C, Denmark\\
$^4$Centre for Astrophysics, University of Southern Queensland, Toowoomba, QLD 4350, Australia\\
$^5$Institute for Astronomy, University of Hawai`i, Honolulu, HI, USA
}

\date{Accepted 16 April 2019.}

\pubyear{2019}

\begin{document}
\label{firstpage}
\pagerange{\pageref{firstpage}--\pageref{lastpage}}
\maketitle

\begin{abstract}
We have used NASA's TESS mission to study catalogued $\delta$ Scuti stars.  We examined TESS light curves for 434 stars, including many for which few previous observations exist.  We found that 62 are not $\delta$~Scuti pulsators, with most instead showing variability from binarity.  For the 372 $\delta$~Scuti stars, we provide a catalogue of the period and amplitude of the dominant pulsation mode.  Using Gaia DR3 parallaxes, we place the stars in the period--luminosity diagram and confirm previous findings that most stars lie on a ridge that corresponds to pulsation in the fundamental radial mode, and that many others fall on a second ridge that is a factor two shorter in period.  This second ridge is seen more clearly than before, thanks to the revised periods and distances.  We demonstrate the value of the period--luminosity diagram in distinguishing $\delta$~Scuti stars from short-period RR~Lyrae stars, and we find several new examples of high-frequency $\delta$~Scuti stars with regular sequences of overtone modes, including XX~Pyx and 29~Cyg.  Finally, we revisit the sample of $\delta$~Scuti stars observed by Kepler and show that they follow a tight period--density relation, with a pulsation constant for the fundamental mode of $Q=0.0315$\,d.

\end{abstract}

% plain text abstrat (for arxiv; 209 words):

%We have used NASA's TESS mission to study catalogued delta Scuti stars.  We examined TESS light curves for 434 stars, including many for which few previous observations exist.  We found that 62 are not delta Scuti pulsators, with most instead showing variability from binarity.  For the 372 delta Scuti stars, we provide a catalogue of the period and amplitude of the dominant pulsation mode.  Using Gaia DR3 parallaxes, we place the stars in the period-luminosity diagram and confirm previous findings that most stars lie on a ridge that corresponds to pulsation in the fundamental radial mode, and that many others fall on a second ridge that is a factor two shorter in period.  This second ridge is seen more clearly than before, thanks to the revised periods and distances.  We demonstrate the value of the period-luminosity diagram in distinguishing delta Scuti stars from short-period RR Lyrae stars, and we find several new examples of high-frequency delta Scuti stars with regular sequences of overtone modes, including XX Pyx and 29 Cyg.  Finally, we revisit the sample of delta Scuti stars observed by Kepler and show that they follow a tight period-density relation, with a pulsation constant for the fundamental mode of Q = 0.0315 d.


%%%%%%%%%%%%%%%%%%%%%%%%%%%%%%%%%%%%%%%%%%%%%%%%%%

%%%%%%%%%%%%%%%%% BODY OF PAPER %%%%%%%%%%%%%%%%%%

\section{Introduction}

High-precision photometry from the \kepler\ and \tess\ space missions is revolutionizing the study of pulsating stars \citep[for recent reviews, see][]{Aerts2021,Kurtz2022}.  Combining this photometry with distances measured with Gaia brings new opportunities to investigate period--luminosity (P--L) relations.
The P--L relations of classical pulsators have a long history, particularly with Cepheids \citep{Leavitt+Pickering1912} and RR\,Lyr variables, making these stars indispensable distance indicators. The P--L relations for their main-sequence cousins---the $\delta$~Scuti stars---were studied by \citet{McNamara1997,McNamara2011} using parallaxes from Hipparcos and have recently been revisited using Gaia parallaxes to refine the luminosities \citep{Ziaali2019,Jayasinghe2020,Poro2021,Gaia-De-Ridder++2022}. However, in many cases the catalogued periods were determined decades ago from ground-based photometry. The \tess\ spacecraft is collecting high-precision all-sky photometry, enabling us to measure pulsation periods more precisely and homogeneously than ever before, without atmospheric variability or large daily gaps in the time series.

We have used \tess\ light curves from the first three years of the mission to study catalogued \dsct\ stars, including many for which few ground-based observations exist, in order to confirm their status and measure their dominant pulsation modes. We use our revised data, together with Gaia DR3 parallaxes, to place these targets more accurately in the period--luminosity diagram.  We build on the work of \citet{Ziaali2019}, who used targets from the heterogeneous \citet{Rodriguez2000} catalogue, and fainter stars observed by the \kepler\ mission, to examine the P--L relation for \dsct\ stars. 

The construction of our sample is described in Sec.~\ref{sec:sample}.  Our procedures for analysing \tess\ data and measuring the dominant pulsation mode of each star are outlined in Sec.~\ref{sec:periods}.
In Sec.~\ref{sec:PL} we plot the P--L relation of our sample and outline its value in distinguishing \dsct\ stars from other pulsators and from non-pulsating stars. We discuss the excess of stars having a dominant pulsation period that is half that of the predicted fundamental mode (Sec.~\ref{sec:second-ridge}), as identified by \citet{Ziaali2019}, and also show the period--density relation for \kepler\ \dsct\ stars (Sec.~\ref{sec:period-density}). In Sec.~\ref{sec:high-frequency}, we discuss stars belonging to the class of high-frequency \dsct\ stars investigated by \citet{Bedding2020}.

\section{Sample Construction}
\label{sec:sample}

Our sample of stars is drawn primarily from the catalogue of \citet[][hereafter R2000]{Rodriguez2000}. In a comprehensive review and compilation of known \dsct\ stars (up to January 2000), R2000 catalogued the primary observational properties of 636 stars, including their dominant pulsation periods and visual magnitudes. We only considered the 407 stars with $V < 12$.
The \tess\ spacecraft observed 301 stars from this list in its first three years (Sectors 1--39).  Of these, 194 were observed at 2-min cadence in at least one sector, and the remaining 107 stars only have full-frame image (FFI) data available, which has 30-minute cadence for Sectors 1--26 and 10-minute cadence from Sector 27 onwards. These \new{301} stars are all included in our sample.

We supplemented our sample with targets from the compilation of 1578 \dsct\ stars by \citet[][hereafter C2013]{Chang2013}, which contains 860 \dsct\ stars not present in R2000. Of these, we analysed the 133 stars with $V < 12$ that have 2-minute \tess\ data available in Sectors 1--39.
Our final sample of 434 stars contains 301 stars from R2000 and 133 additional stars from C2003.

\section{Pulsation periods}
\label{sec:periods}

\subsection{\tess\ light curves}

We queried the \tess\ data products available on the Mikulski Archive for Space Telescopes (MAST)\footnote{\url{https://archive.stsci.edu/}} for each star using the \textsc{Python} \textsc{Lightkurve} package \citep{lightkurvecollaboration2018}. We downloaded light curves for 2-min targets, whereas for targets with only FFI data we used the \textsc{Lightkurve} wrapper \textsc{Platypus}\footnote{\url{https://github.com/danhey/platypus}} to extract and correct the time-series data. In both cases, we used all available data from \tess\ Sectors 1--39.

We computed the Fourier transform of each light curve in order to measure the dominant pulsation modes. Due to the non-zero \tess\ integration times, the measured amplitudes are smaller than the intrinsic amplitudes. To account for this averaging, sometimes referred to as `smearing' or `apodization', we divided the measured Fourier amplitudes by the function $\sinc \left({\pi}{\nu}{t_{\rm int}}\right)$, where $t_{\rm int}$ is the integration time. Because there is no dead time between integrations, 
%(as there is in ground-based observations,)
this equation can be written as $\sinc \left(\frac{\pi}{2} \frac{\nu}{\nu_{\rm Nyq}}\right)$, where $\nu_{\rm Nyq}$ is the Nyquist frequency \citep[e.g.,][]{Huber2010, murphy2012}. We note that this correction is more important for the light curves made with 30-min FFI data. 

A simple peak-finding algorithm was used to extract the peak with the largest corrected amplitude (i.e. the dominant mode) for each star. This was adequate for our goals, given our focus is identifying \dsct\ stars and measuring the dominant modes, rather than carrying out a full frequency analysis on every star. We limited the search to peaks above 3~\cd\ to capture pressure mode (p-mode) pulsations. Below this limit, pulsation peaks are typically gravity or Rossby modes \citep[e.g.,][]{Li++2020}. 

The period--luminosity relation outlined in Section 3 was also used as a guide to confirm that we were recording the correct pulsation modes. Stars that fell significantly to the right of the P--L relation (that is, those with periods too long to be \dsct\ pulsations) were inspected to confirm their variability. A number of these stars were identified as binaries showing no evidence of \dsct\ pulsations. Other stars were able to be amended with the correct \dsct\ pulsation mode. For example, the highest peak in the spectrum of some stars (e.g., BR Hyi and V353 Tel) was in fact due to binarity, but there were sometimes clear \dsct\ pulsations and so the dominant \dsct\ mode was manually recorded. These results are compiled in Table~\ref{tab:freq_table}, which lists the dominant pulsation mode that we have measured in each star from the \tess\ data, along with the corrected amplitude. 

\begin{figure}
\includegraphics[width=1.0\linewidth]{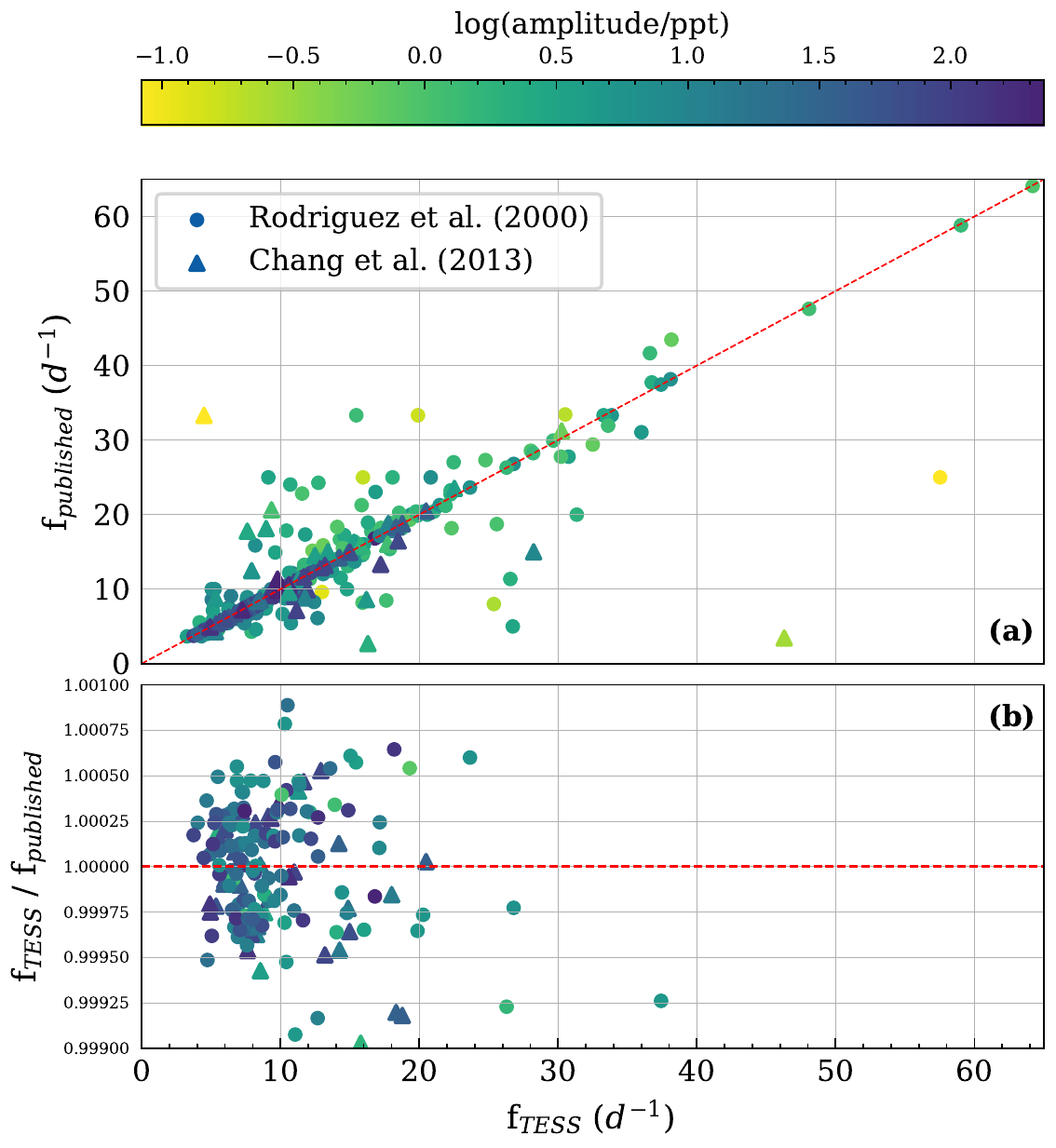}
\caption{{\bf (a)} Comparison between the catalogued frequencies and those measured from \tess. The red dotted line shows a ratio of 1:1. Stars included in the R2000 catalogue are plotted as circles, while those added from the C2013 catalogue are plotted as triangles. {\bf (b)} Close-up of the ratios of values, highlighting stars whose \tess-revised dominant modes are close to their previously published values. \new{The points indicate the dominant pulsation mode in each star and are colour-coded by the amplitude (on a logarithmic scale).}}
\label{fig:ratio_plot}
\end{figure}

\subsection{Non-\texorpdfstring{$\delta$}{delta}~Scuti stars}
% \texorpdfstrong prevents a LaTeX warning when maths is used in a section heading
\begin{table}
    \centering
    \caption{Compilation of 63 stars identified as non-\dsct\ stars. Column R2000 is set to 1 for stars listed in \citet{Rodriguez2000}, while the remainder are from \citet{Chang2013}.}
    \input{removed_star_sample}
    \label{tab:removed_targets_table}
\end{table}

By inspecting the \tess\ light curves and consulting publications since R2000, we identified 62 stars from our sample that showed no evidence of \dsct\ pulsations (see Table~\ref{tab:removed_targets_table}). Of these, 21 stars were included in the original R2000 catalogue and 41 were added from C2013. %We give details on these stars below.

\citet{Ziaali2019} already flagged the following eight stars from R2000 as not having \dsct\ pulsations, which we confirmed with \tess\ data:
\begin{itemize}
    \item V753~Cen, TV~Lyn, and UY~Cam are RR Lyrae variables \citep[e.g.,][] {McNamara2011,Sneden2018};
    \item BQ~Phe, DE~Oct, and V345~Gem are binaries with no pulsating component \citep{Liakos+Niarchos2017};
    \item AK~Men was also found to be a binary with no pulsating component; and
    \item V1228~Cen is a B-type $\beta$\,Cep variable \citep{Pigulski2008}.
\end{itemize}
%We identified an additional seven stars from the R2000 catalogue that are not \dsct\ variables. 
Of the 41 stars removed from C2013, several are in binary systems with no pulsating component. Their Fourier spectra often look similar to those of high-amplitude $\delta$\,Sct stars (HADS) and were initially flagged for inspection after placement in the P--L diagram, for falling outside of the expected area. Finally, HD~229085 shows p-mode pulsations but is too luminous to be a \dsct\ star ($M_V = -0.55$); being only 1.0\,deg from the Galactic plane, it appears to be a heavily-reddened $\beta$~Cephei star. Overall, we confirm the value of the P--L relation in helping to identify non-\dsct\ pulsators that we might otherwise misclassify. 

\subsection{Comparison with published periods}

Figure~\ref{fig:ratio_plot} compares the dominant pulsation mode recorded in the original catalogues with our values from \tess. The values agree well for the majority of our sample: periods for 67 percent of the stars are within 10\% of the original value. 
It appears that stars with larger differences are primarily those with amplitudes smaller than 1\,mmag. This is presumably a result of lower signal-to-noise ratio and poorer time coverage of the ground-based observations, making it difficult to accurately identify and measure the dominant mode. Some \dsct\ stars are also known to undergo amplitude modulation \citep[see][and references therein]{Bowman2016}, which may have caused a new mode to become dominant since the previous studies. 

%Figure~\ref{fig:ratio_plot} (b) shows that stars with higher amplitudes (i.e. with amplitudes greater than 10\,mmag) are more consistent between the ground-based observations and the updated \tess\ data. This is in agreement with the suggestion that the lower-amplitude stars had greater error in their original measurements. In addition, Figure~\ref{fig:ratio_plot} (b) highlights the value of the greater precision of the \tess\ measurements: lower amplitude stars are now able to be placed more correctly in the P--L diagram.

\section{Period--Luminosity Relation}
\label{sec:PL}

Revising the dominant modes of our sample allows them to be placed more accurately in the period--luminosity diagram. We calculated the absolute $V$-band magnitudes ($M_V$) for these stars using:
\begin{equation} \label{abs_mag_rod}
M_{V} = V + 5\log_{10}\pi + 5 - A_V,
\end{equation}
where $\pi$~is the parallax in arcsec. We  used parallaxes from {\em Gaia} DR3, except in the cases where the parallax from {\em Hipparcos} was more precise. The extinction due to interstellar dust ($A_V$) was calculated by fitting the spectral energy distribution (SED) of each star using the \textsc{astroARIADNE}\footnote{\url{https://github.com/jvines/astroARIADNE}} \textsc{python} package. For each star, we queried photometric data from the \textsc{Vizier} database \citep{vizier}. The stars were then fitted to the \citet{castelli&kurucz2003} model using a nested sampling algorithm ({\sc dynesty}, \citealt{speagle2020}) to obtain extinctions. In order to obtain a uniform sample of apparent ($V$) magnitudes, we used Tycho $V_T$ and $B_T$ magnitudes converted to the Johnson $V$ photometric system using the linear transformation
\begin{equation} 
\label{eq:tycho_conversion}
V = V_{T} - 0.090(B_{T} - V_{T}),
\end{equation}
for which systematic errors do not exceed 0.015 magnitudes \citep{ESA1997}. In plotting, we further limited our sample to stars with fractional parallax uncertainties < 5\%, and extinction $A_{\rm V} < 0.2$.

\begin{figure*}
\includegraphics[width=0.7\linewidth]{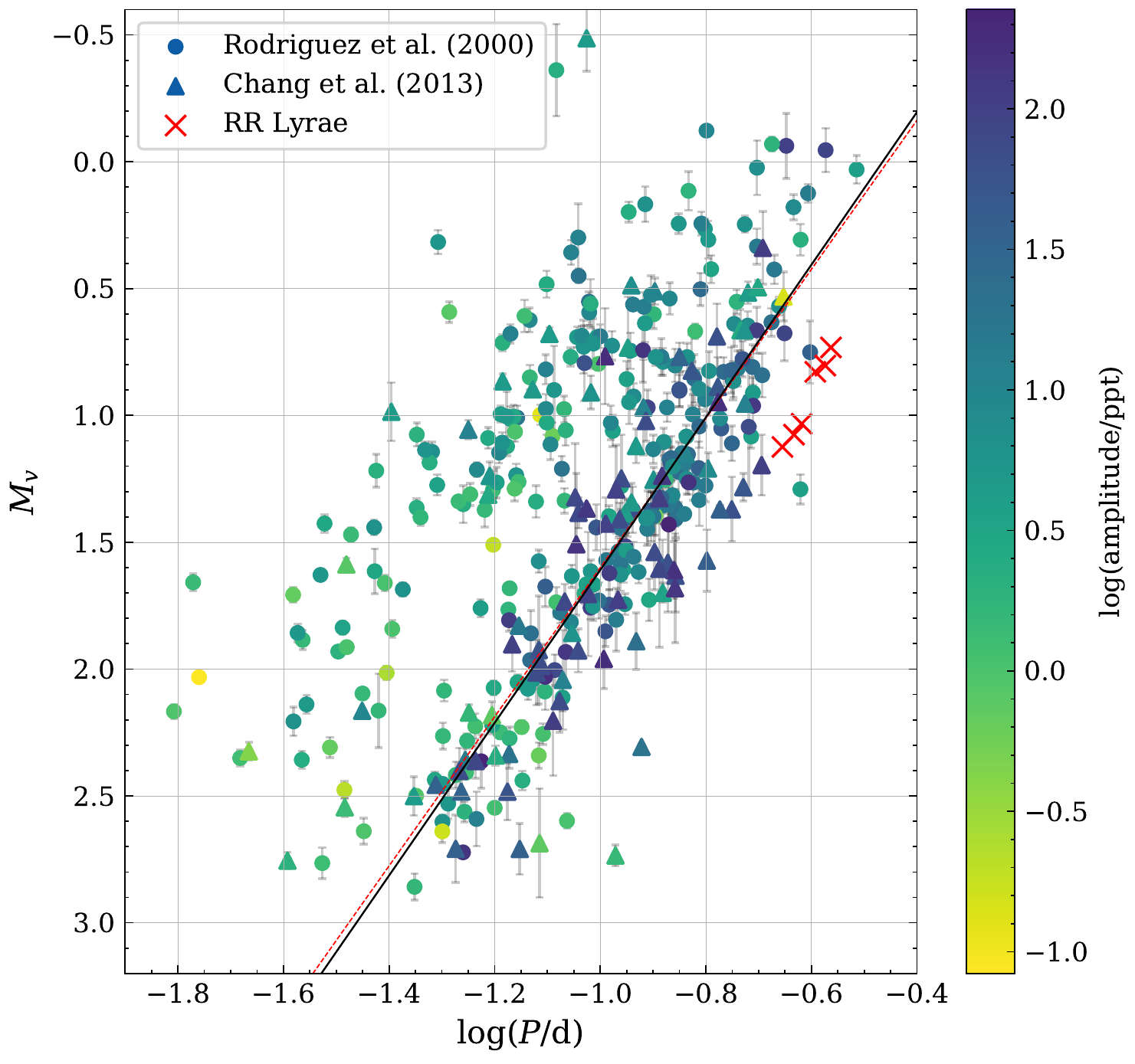}
\caption{Period--luminosity diagram of \dsct\ sample, using periods from \tess\ data and distances from Gaia DR3.  \new{The points indicate the dominant pulsation mode in each star and are colour-coded by amplitude (on a logarithmic scale).} The stars plotted are restricted to variables having apparent $V$ magnitudes brighter than 12.0, and fractional parallax uncertainties less than 5 percent. The dashed red line is the P--L relation fitted by \citet{Ziaali2019} and the solid black line is our revised relation (Eq.~\ref{eq:newfit}).}
\label{fig:PL_revised}
\end{figure*}

Figure~\ref{fig:PL_revised} shows our updated period--luminosity relation, colour-coded by pulsation amplitude. The red crosses show RR~Lyrae variables, which have their own P--L relation \citep{McNamara2011, Sneden2018, Molnar++2022, Garofalo++2022}.  Indeed, when accurate parallaxes are available, the P--L diagram is a useful way to distinguish between HADS and short-period RR\,Lyr variables (so-called RRc stars, which pulsate in the first overtone), whose light curves look very similar. 
The dashed red line in Fig.~\ref{fig:PL_revised} is the P--L relation fitted by \citet{Ziaali2019}, 
\begin{equation} \label{eq:ziaalifit}
M_V = (-2.94 \pm 0.06)\log(P/{\rm d}) - (1.34 \pm 0.06),
\end{equation}
which was based on periods from R2000 and parallaxes from Gaia DR2.  This result is similar to the P--L relation derived by \citet{McNamara2011}, which was fitted to HADS stars thought to pulsate in their fundamental radial mode. We have performed a new fit using our revised measurements, which gives a slightly steeper P--L relation:
\begin{equation} \label{eq:newfit}
M_V = (-3.01 \pm 0.07)\log(P/{\rm d}) - (1.40 \pm 0.07).
\end{equation}
This updated relation is show as the solid black line in Fig.~\ref{fig:PL_revised}.

%HADS tend to oscillate in fundamental radial mode? so this makes sense? 
We see in Fig.\,\ref{fig:PL_revised} that higher-amplitude stars tend to fall closer to the fundamental-mode ridge, while lower-amplitude stars show more scatter. This main ridge is also sharper than that of \citet{Ziaali2019}, and there are fewer stars falling to the lower right of the relation. Using our revised periods and distances has also made the second ridge, lying to the left of the fundamental-mode ridge, more distinct than before.

%note: vertical displacement from the Ziaali line... originally attributed to not including extinction values but we note that it seems like the original P--L diagram had the same vertical offset... unsure where it's coming from

\begin{figure}
\includegraphics[width=1.0\linewidth]{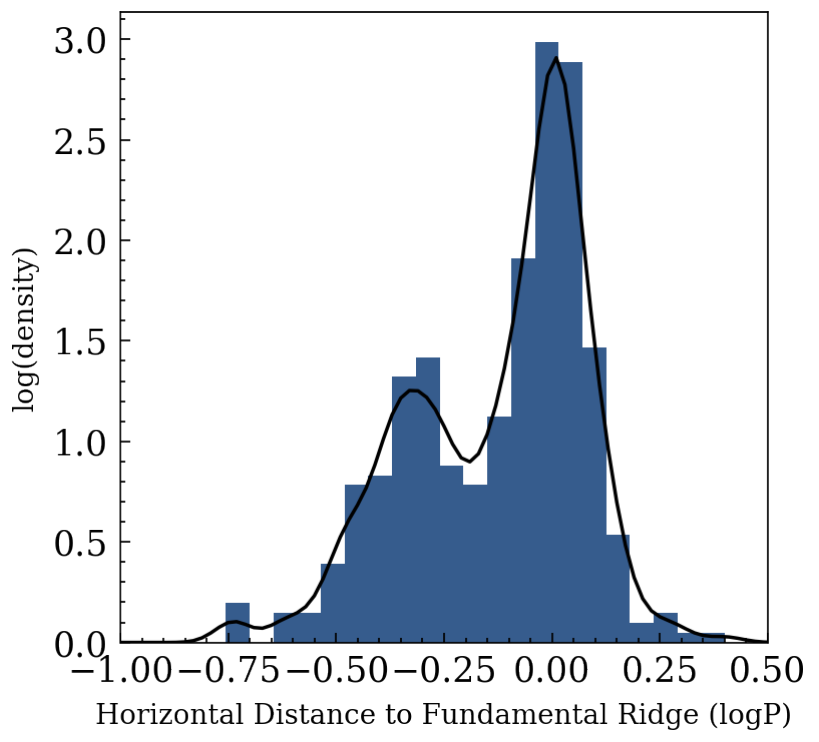}
\caption{Histogram showing the horizontal distance of data points to the P--L relation in Eq.~\ref{eq:newfit}, overlaid with a fixed-bandwidth kernel density estimate (KDE) highlighting the same excess of stars displaced by $-0.3$ in $\log P$ from the fundamental mode ridge.}
\label{fig:histogram_updated}
\end{figure}  

Figure~\ref{fig:histogram_updated} shows the horizontal distance of data points to Eq.~\ref{eq:newfit} as a histogram. We confirm a strong excess of stars displaced to the left by 0.30 in $\log P$, corresponding to a period ratio of 0.50. Since histogram results can depend on bin size, we also added a fixed-bandwidth kernel density estimate (KDE; \citealt{Terrell+Scott1992}) for the same data to Fig.\,\ref{fig:histogram_updated}. The excess of stars displaced to the left of 0.3 in $\log P$ is still present in the KDE, lending support that the excess seen in the histogram is real. 

\subsection{Second Ridge Stars}
\label{sec:second-ridge}

The excess of second-ridge stars was identified by \citet{Ziaali2019}, \new{and also noted by \citet{Jayasinghe2020} and \citet{Poro2021}. Note that \citet[][their figure~4]{Poro2021} identified this ridge with first-overtone modes but that is not consistent with the well-established period ratio of 0.77 between the fundamental and first overtone in \dsct\ stars \citep{Petersen+CD1996}. Rather, the periods on the second ridge coincide with the third or fourth overtones \citep{Ziaali2019,Jayasinghe2020}, but the reason for the excess of stars with these periods remains to be explained.}  

\begin{figure*}
\includegraphics[width=1.0\linewidth]{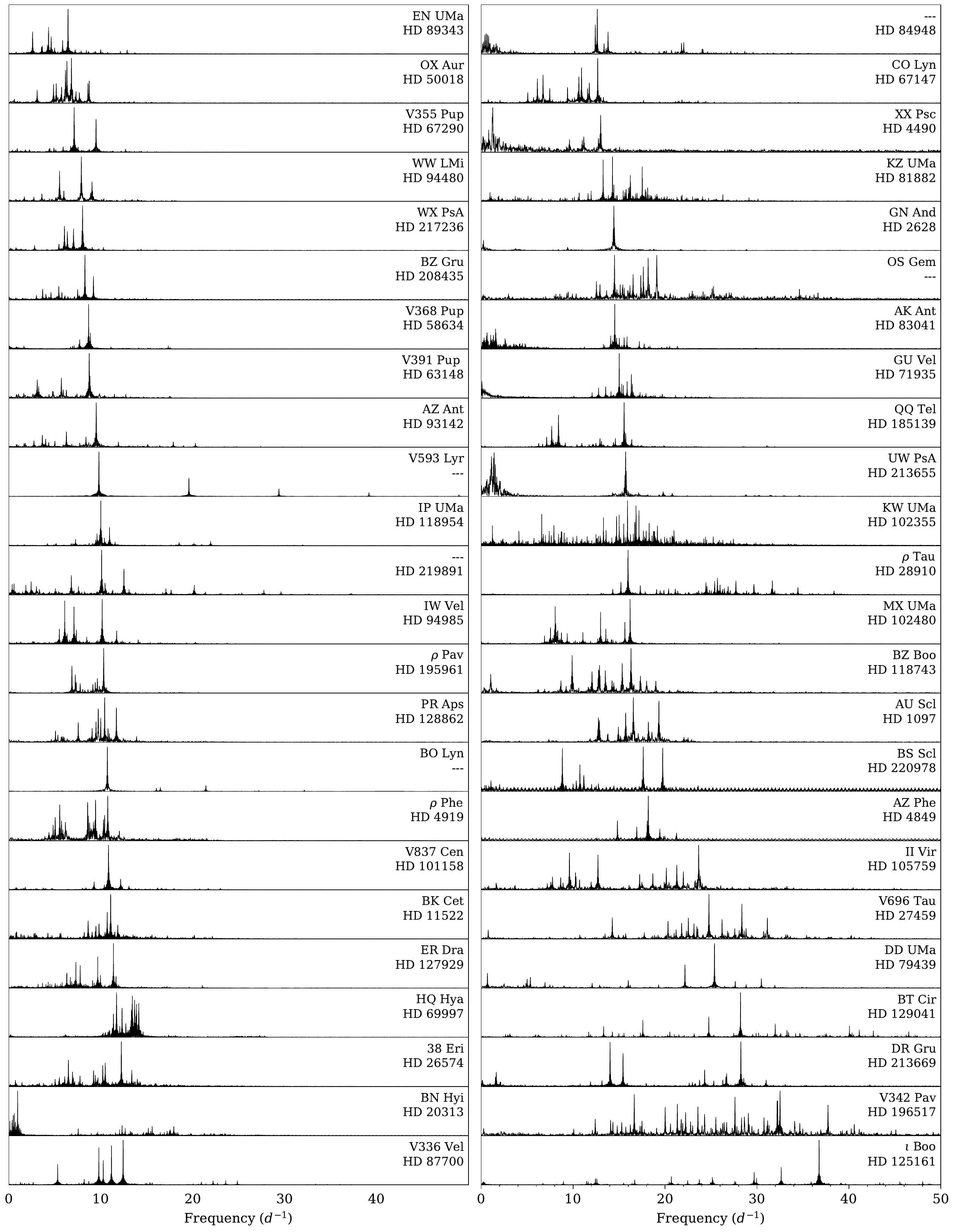}
\caption{Pulsation spectra of the 48 \dsct\ stars lying on the second ridge of our Period--Luminosity diagram, ordered by increasing frequency of the dominant pulsation mode. The vertical axis is amplitude in normalized units.}
\label{fig:second_ridge_stars}
\end{figure*}

Figure~\ref{fig:second_ridge_stars} shows the amplitude spectra of the 48 stars whose horizontal displacement in $\log P$ from the main ridge falls in the range $-0.35$ to $-0.25$. In many cases, we can see a peak at half the frequency of the highest peaks, corresponding to the expected position of the fundamental mode. Conversely, some stars appear to only coincidentally lie on the second ridge. These are typical \dsct\ and even HADS that are displaced, perhaps due to binarity or some other error in their absolute magnitudes. 

We note, for example, the following stars that appear to be shifted vertically in absolute magnitude due to binarity:
\begin{itemize}

    \item BO~Lyn looks like a typical HADS with main peak at 10.71 (60 ppt), plus weaker peaks at 16.47 (5 ppt) and 16.04 (4 ppt). We suspected it to be a binary, and therefore vertically displaced in absolute magnitude, as the dominant mode of HADS should fall on the fundamental-mode P--L sequence. Indeed, \citet{Li2018} suggested it has an A-type companion in a 35-yr orbit, based on variations in the pulsation phase over the past century.
    
    \item $\rho$~Tau (HD~28910) is a member of the Hyades, with a Hipparcos parallax of 20.61 mas which matches the cluster.  It lies above the cluster isochrone in the colour--magnitude diagram, and \citet{Antonello+Pasinetti-Fracassini1998} suggest it is a binary.  Hence, it is probably shifted up and not a true second-ridge star.
    
    \item RS~Cha (9~Cha; HD~75747; not shown in Fig.~\ref{fig:second_ridge_stars} because its horizontal displacement is $-0.23$) is a contact EB consisting of two pre-main-sequence stars with tidally perturbed pulsations \citep{steindletal2021}.  Since these stars have similar luminosities, our absolute magnitude is too bright by about 0.7 and has shifted the star towards the second ridge.
    
\end{itemize}

These exceptions are few: the majority of stars on the second ridge are not binaries. Furthermore, we confirm the finding by \citet{Ziaali2019} that known binary systems do not lie preferentially on the second ridge. 

An alternative explanation may come from amplitude modulation and mode resonances.
We have discussed amplitude modulation on the timescales of decades as contributing to changes between the measurements from our source catalogues (R2000 or C2013), and \tess\ measurements of the dominant pulsation mode. While the timescales of continuous \tess\ observations of our sample are too small to conduct a complete analysis of amplitude modulation, a number of stars in our sample have \tess\ observations separated by a year or more. Interestingly, of these, several second ridge stars show modulation of the second-ridge mode between observations, so that the fundamental radial mode becomes the dominant mode. 

%TimB: Interesting question: do second-ridge stars have a higher incidence of amplitude modulation than the general population?

\begin{figure}
\includegraphics[width=1.0\linewidth]{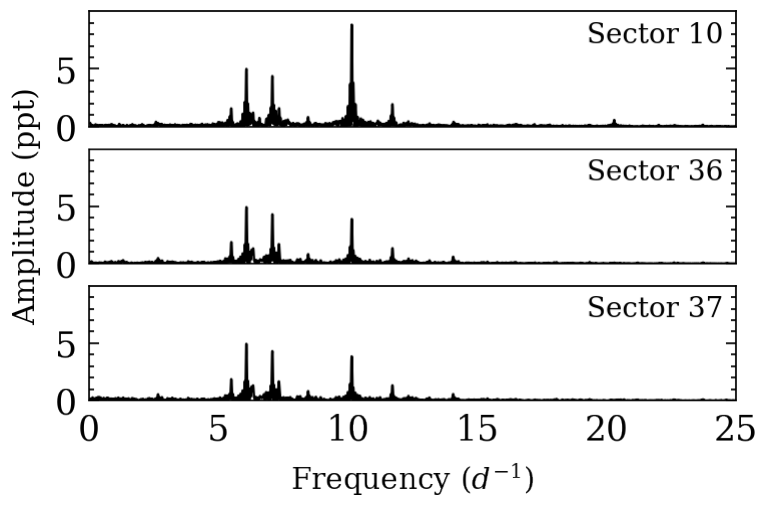}
\caption{Pulsation spectra of IW Vel, a second-ridge star that shows changes in amplitude over time (see Sec.~\ref{sec:second-ridge}).}
\label{fig:iw_vel}
\end{figure}

IW~Vel (HD~94985) is one such example. IW~Vel was observed by \tess\ in Sector 10 in 2019, and again in Sectors 36 and 37 in 2021. Figure~\ref{fig:iw_vel} shows clear amplitude modulation of the second ridge mode (10.15\,\cd) between Sectors 10 and 36, two years apart, continuing to  Sector 37. In total, the amplitude drops by 67 percent between Sectors 10 and 37, with a 25 percent decrease occurring between Sectors 36 and 37. This moves IW~Vel off the second ridge.
%This decrease is accompanied by a small increase in the amplitude of frequency at 6.08\cd ({double check predicted fund. mode, might be the peak to the left}), which increases by 1 percent in total, such that it is the dominant frequency in Sectors 36 and 37. 

Similarly, BS~Scl was observed by \tess\ in Sector 2 in 2018, and again in Sector 29 in 2020. In Sector 2, the second ridge mode at 17.63\,\cd\ is dominant with an amplitude of 1.38\,ppt, while the fundamental radial mode peak at 8.83\,\cd\ has an amplitude of 1.29\,ppt. By Sector 29 however, the second ridge mode has decreased in amplitude by 27\:percent and is no longer the dominant frequency. 
%Meanwhile the amplitude of the fundamental mode increased by about 4 percent. 
We also note that the peak at 11.2\,\cd\ undergoes significant modulation, decreasing in amplitude by about 42 percent.

In summary, it is possible that stars occupy the second ridge only temporarily, perhaps when a mode of higher radial order is boosted to a higher amplitude through interactions with other modes.

Finally, we note that the amplitude spectrum of HQ~Hya (Fig.~\ref{fig:second_ridge_stars}) is similar to those of a new class of \dsct\ stars that have been identified in \textit{Kepler} data (see Fig.~B12 of \citealt{Murphy2019}).  These stars are characterised by a broad power excess containing many closely spaced peaks, rather than the more widely separated peaks that are seen in most \dsct\ stars.  We are not aware of an explanation of this phenomenon but we have noticed several other examples in our sample (e.g., BG~Hyi, AI~Scl, VV~Ari and 60~Tau = V775~Tau).

%question of amplitude modulation? otherwise highlighting those second ridge stars, separated into groups (including stars on the ridge coincidentally)
%interesting to follow up Rodriguez second ridge observations with those of Kepler stars --> larger sample, further examination of reasons behind excitation of modes double the frequency of the fundamental

% moved this section to here
\subsection{Period--density relation of the {\em Kepler} \texorpdfstring{$\delta$}{delta}~Scuti sample}
\label{sec:period-density}
%in comparison with stars from Murphy+ 2019 -- histogram from line only, now using Gaia DR3, confirms presence of second ridge in sample (although not as sharp, still clearly there)

There have been numerous studies of the several thousand \dsct\ stars observed by the \kepler\ Mission (for reviews, see \citealt{Guzik2021,Kurtz2022}).  Here, we take advantage of the Gaia parallaxes to investigate their period--density relation.
There is a well-established relationship between the period of a pulsating star and its mean density \citep[e.g.,][sec.~5.2]{Catelan+Smith2015}.  The relation is usually written as
\begin{equation}
    P = Q \left(\frac{\rho}{\rho_\odot}\right)^{-0.5}, \label{eq:period-density}
\end{equation}
where the pulsation constant, $Q$, has a different value for each mode.\footnote{Note that Eq.~\ref{eq:period-density} applies to pressure modes (p~modes), which include pulsations in $\delta$~Scuti stars, $\beta$~Cephei stars and solar-like oscillators.  It does not apply to gravity modes (g~modes), such as those in $\gamma$~Doradus stars, Slowly Pulsating B stars and white dwarfs.}
For the fundamental radial mode in \dsct\ stars, $Q$ is about 0.033\,d \citep[e.g.,][]{Fitch1981,North1997,Breger2000,Lovekin+Guzik2017}.

\begin{figure}
\includegraphics[width=1.0\linewidth]{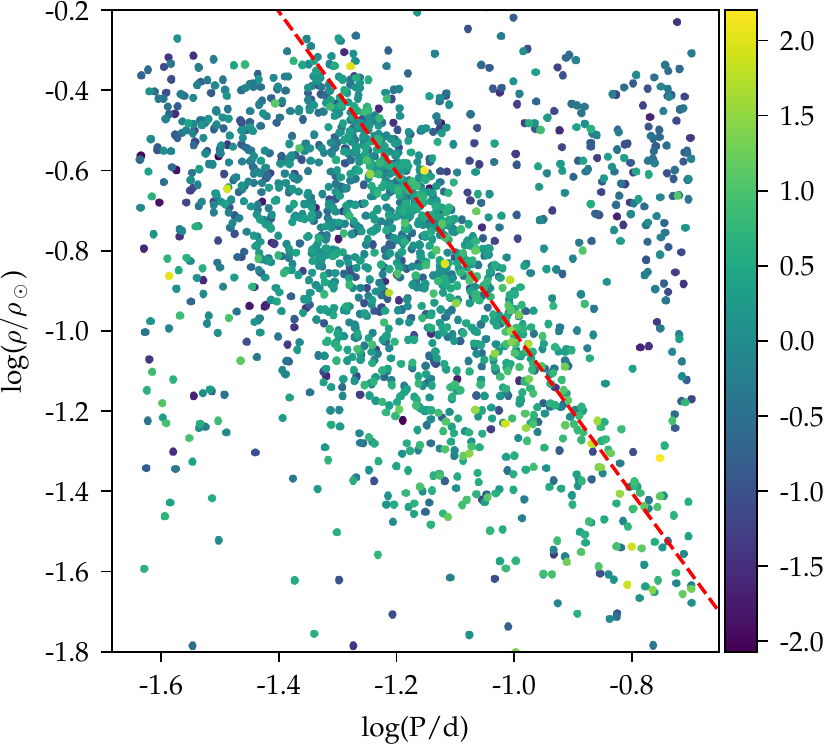} % line is Q=0.0315
\caption{Period--density relation for \kepler\ \dsct\ stars.  Points show the dominant pulsation mode in each star, colour-coded by the logarithm of the pulsation amplitude in ppt.  Pulsation periods and amplitudes were taken from \citet{Murphy2019} and densities were calculated from the {\em Kepler Stellar Properties Catalog} \citep{Berger2020}.  The red dashed line shows the theoretical period--density relation (Eq.~\ref{eq:period-density}) with a pulsation constant of $Q=0.0315$\,d.  Points in the upper right are \gdor\ stars \citep[see][]{Ziaali2019}.}
\label{fig:period_density_kepler}
\end{figure}

Calculating densities for a large number of stars requires estimating masses by fitting evolutionary tracks in the H--R diagram.  This has been done using Gaia DR2 parallaxes for the full set of \kepler\ targets by \citet{Berger2020} and we used their masses and radii to estimate densities the \kepler\ \dsct\ stars.
Figure~\ref{fig:period_density_kepler} show the period--density relation, where period of the dominant mode in each star was taken from \citet{Murphy2019}.  The diagram looks very similar to the period--luminosity relation \citep{Ziaali2019} but with the vertical axis reversed, which is expected because of the tight inverse correlation between luminosity and density.  Most points in Fig.~\ref{fig:period_density_kepler} lie on a ridge that corresponds to the radial fundamental mode.  This is confirmed by the dashed line, which shows the theoretical relation (Eq.~\ref{eq:period-density}) with $Q=0.0315$\,d.  We also see clearly the excess of points on the second ridge, with half this period. \new{Note that points in the upper-right corner of Fig.~\ref{fig:period_density_kepler} are \gdor\ pulsators, as discussed by \citet{Ziaali2019}.}

%%%%%%%%%%%%%%%%%%%%%%%%%%%%%%%%%%%%%%%%%%%%%%%%%%%%%
\section{High-frequency \texorpdfstring{$\delta$}{delta}~Scuti stars}
\label{sec:high-frequency}

Some of the stars in our sample belong to the class of high-frequency \dsct\ stars, discovered by \citet{Bedding2020} to have remarkably regular sequences of overtone modes.  Three were listed in that paper, namely EX~Eri (HD~30422), V435~Car (HD~44958) and V349~Pup (HD~59594), but only for V435~Car was the \echelle\ diagram shown.  The \echelle\ diagram for EX~Eri was shown by \citet{Murphy2020}, who also showed \echelle\ diagrams for four other stars in our sample: V1023~Cen (HD~102541), MO~Hya (HD~111786), V346~Pav (HD~168740) and HD~210111. We show two more examples, XX~Pyx and 29~Cyg, in the next sections. It is noteworthy that in many high-frequency \dsct\ stars, the fundamental mode is not the strongest mode \citep{Bedding2020}. Such is the case for all five Pleiades $\delta$\,Sct stars with \kepler-K2 data that were recently studied by \citet{murphy2022}.

\subsection{XX~Pyx}

\begin{figure}
\centerline{\includegraphics[width=\linewidth]{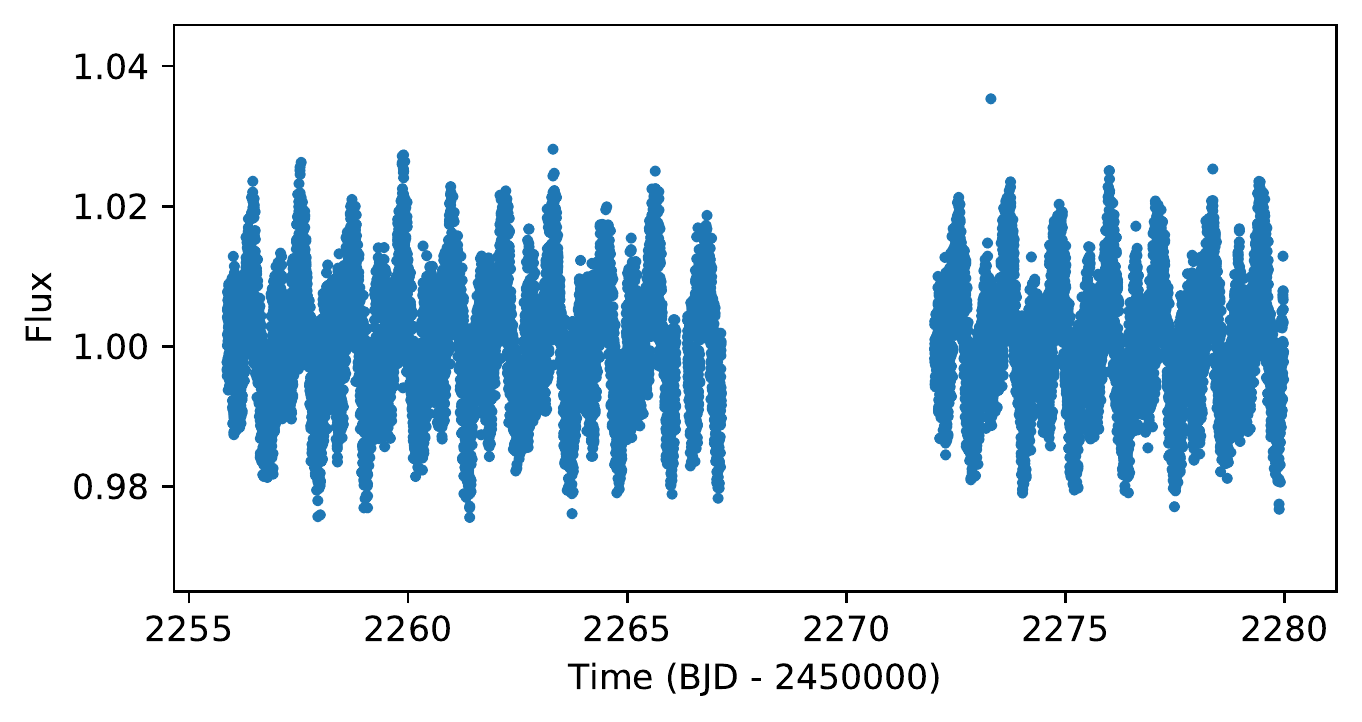}}
\centerline{\includegraphics[width=\linewidth]{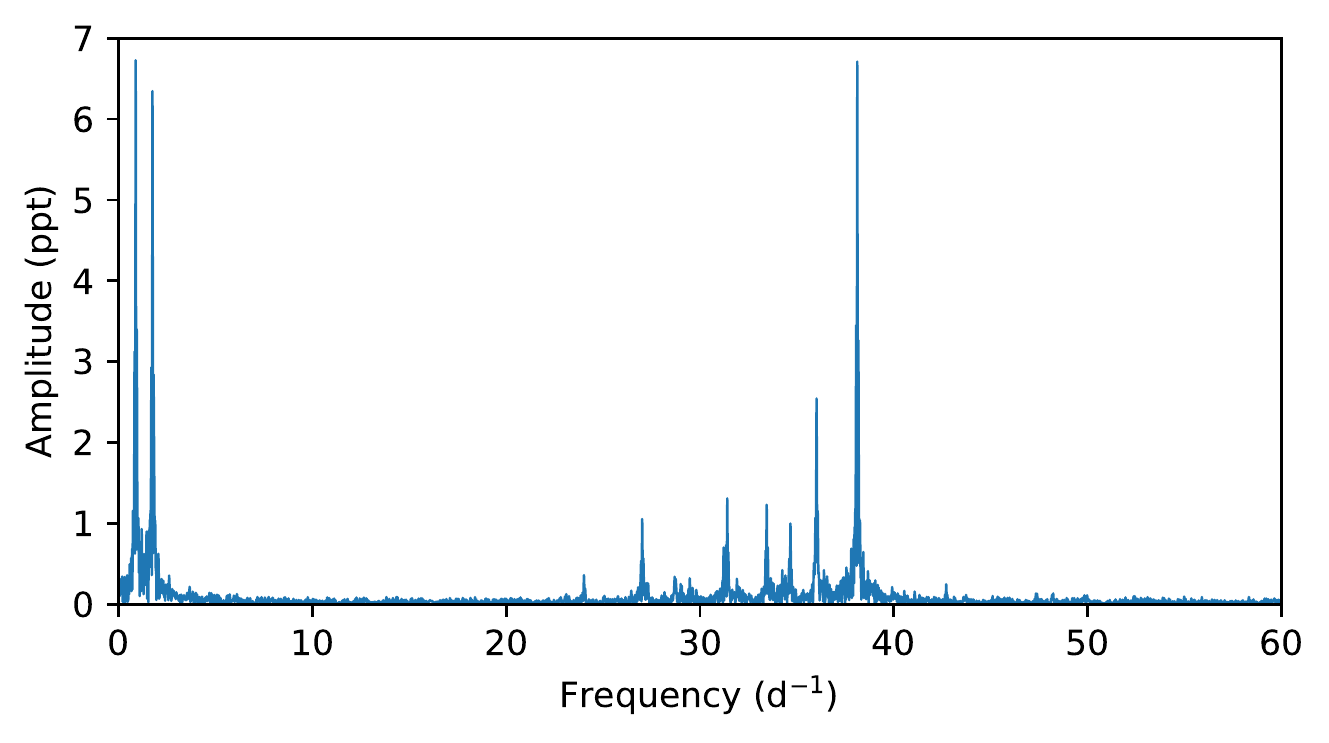}}
\centerline{\includegraphics[width=0.8\linewidth]{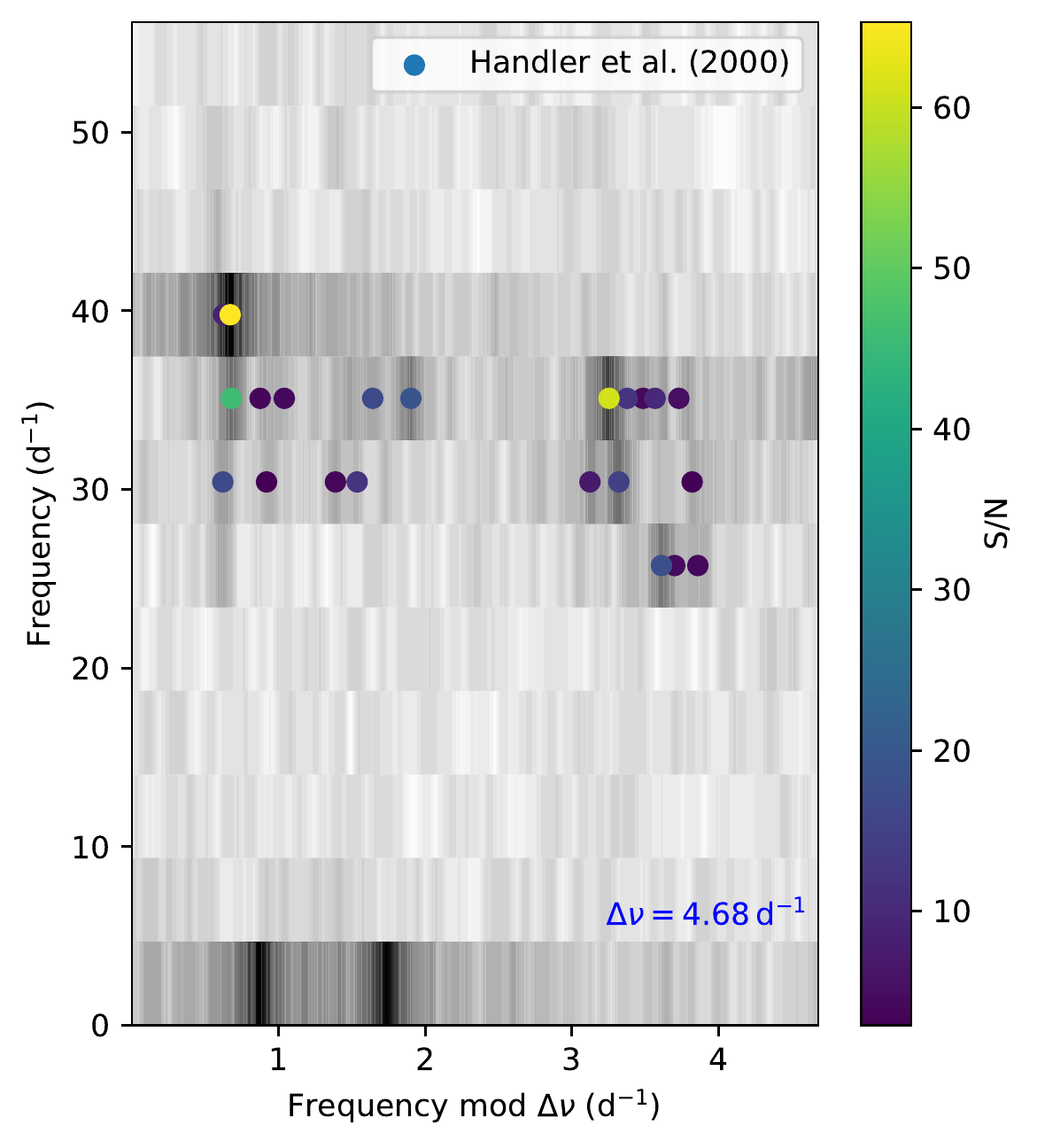}}
\caption{\tess\ observations of XX~Pyx (Sector~35), showing the light curve (top), the amplitude spectrum (middle), and the amplitude spectrum in \echelle\ format.  The circles in the \echelle\ diagram show the frequencies reported by \citet{Handler++2000}, colour-coded by signal-to-noise (their Table~4).}
\label{fig:xx_pyx}
\end{figure}

XX~Pyx is a multi-periodic pulsator that has been the subject of three ground-based multi-site campaigns \citep[][and references therein]{Handler++2000}.
The star was not observed by \tess\ in Cycle~1 because it fell in a gap between fields, but it was observed in Cycle~3 (Sector~35).  The \tess\ light curve (Fig.~\ref{fig:xx_pyx}) confirms that it is an ellipsoidal variable with a binary period of 1.15\,d, as found by \citet{Aerts2002}.

\citet{Handler++2000} detected 19 independent pulsation frequencies in the range 27--38\,\cd, and suggested a large separation of 4.63\,\cd.  By generating an \echelle\ diagram with those frequencies, \citet{Bedding2020} confirmed that a value of $\Dnu = 4.70\,\cd$ gave a reasonably good alignment of the peaks.  Figure~\ref{fig:xx_pyx} shows the \tess\ amplitude spectrum in \echelle\ format.  We confirm that $\Dnu = 4.68\,\cd$ produces a ridge on the left-hand-side that we can identify as $l=1$ modes (see \citealt{Bedding2020} for discussion of similar stars).  Three of these modes were detected by \citet{Handler++2000}, and we see two additional weaker $l=1$ modes in this ridge.  This star is clearly worthy of further study.

\subsection{29 Cyg}

\begin{figure}
\includegraphics[width=1.0\linewidth]{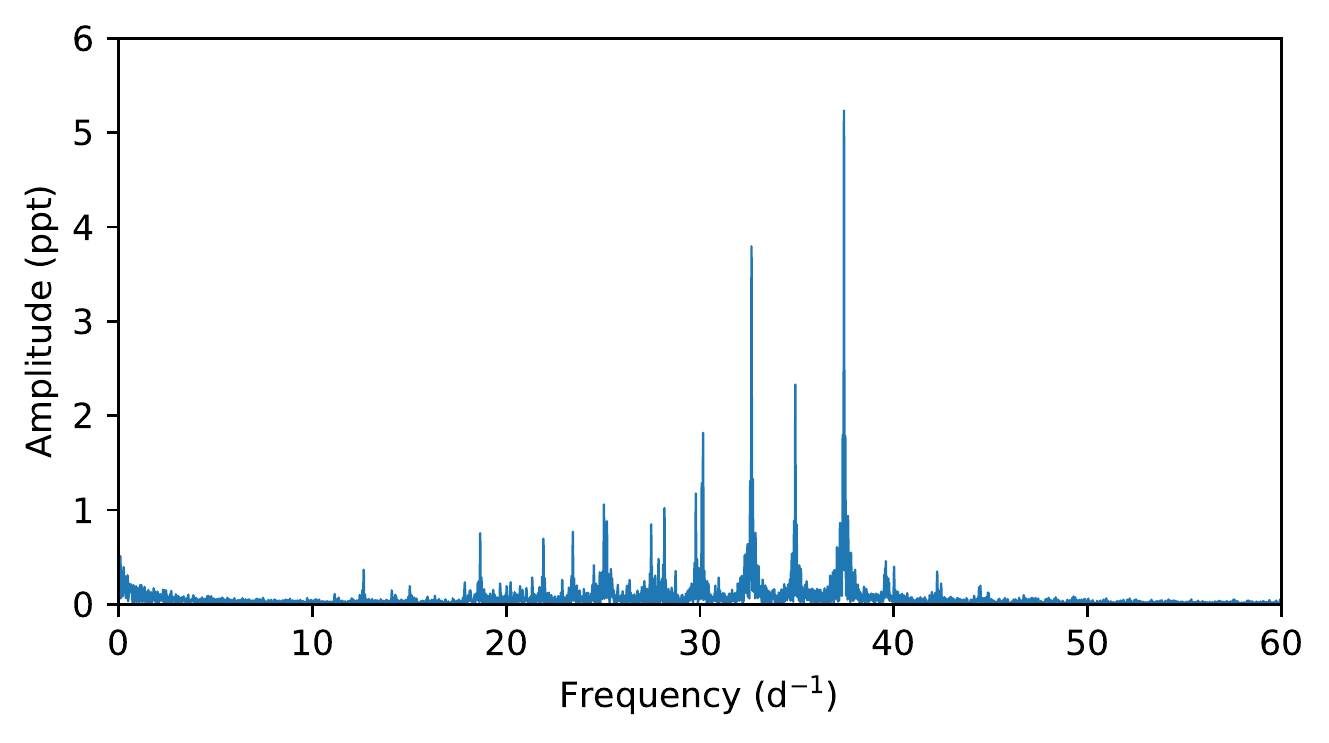}
\centerline{\includegraphics[width=0.8\linewidth]{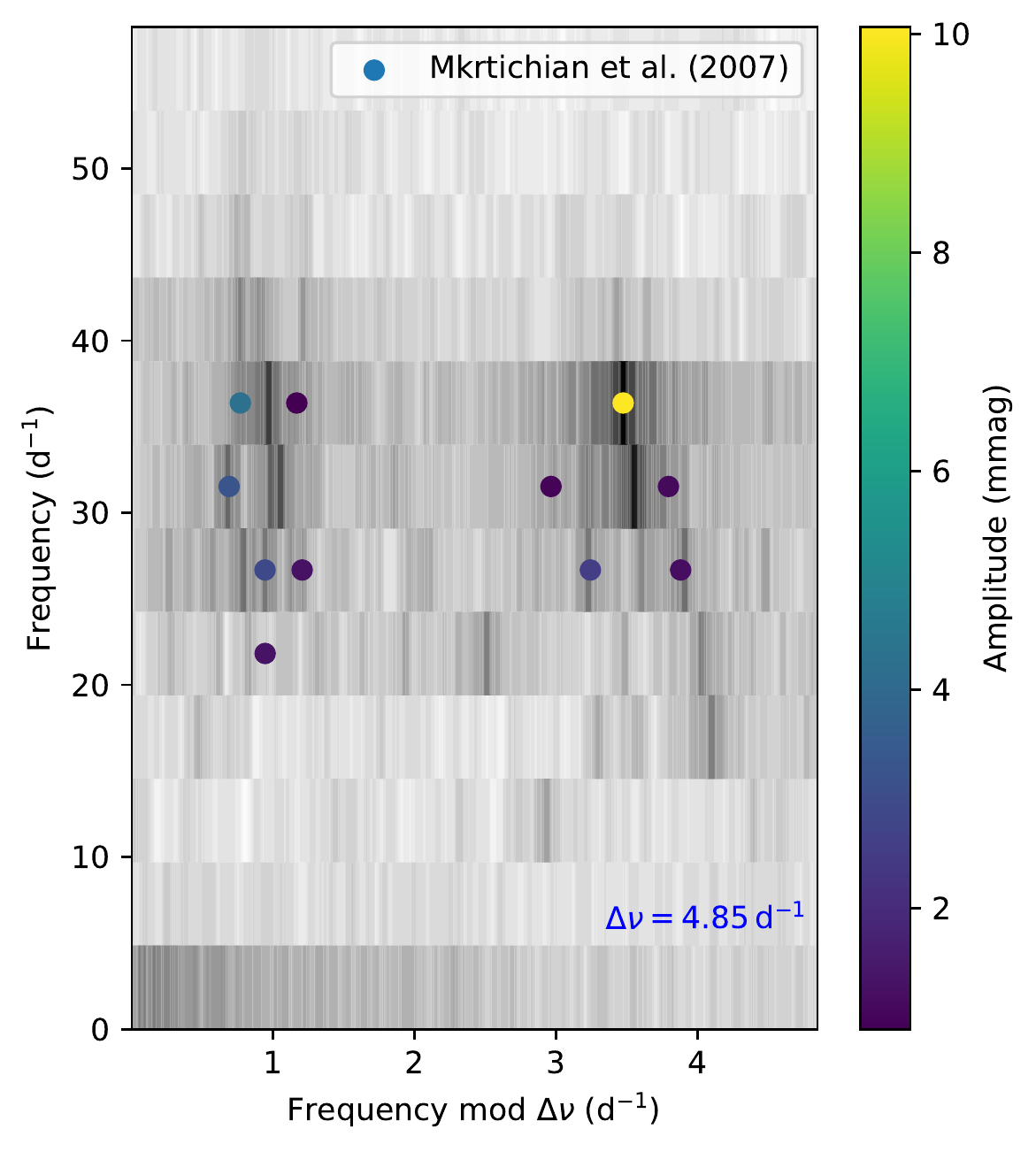}}
\caption{\tess\ observations of 29~Cyg (Sectors~14 and 15), showing the amplitude spectrum (top), and the \echelle\ diagram (bottom).  The circles show the frequencies reported by \citet{Mkrtichian2007}, colour-coded by amplitude (their Table~8).}
\label{fig:29_cyg}
\end{figure}

29 Cyg (V1644~Cyg; HD~192640) is a multi-periodic pulsator that has been the subject of two ground-based multi-site campaigns by \citet{Mkrtichian2007}.  They measured 11 modes in the range 20--37\,\cd\ and suggested a large separation of $\Dnu = 4.82$\,\cd.
Theoretical models using these frequencies were calculated by \citet{Casas2009}. The \tess\ data, shown in Fig.~\ref{fig:29_cyg}, confirm that 29~Cyg is a high-frequency \dsct\ star with a large separation of $\Dnu=4.85$\,\cd.  With a magnitude of $V=4.9$, this makes 29~Cyg the brightest known member of this class of stars and an ideal target for follow-up observations. It is also known to be a $\lambda$\,Boo star, which is common among high-frequency $\delta$\,Sct stars \citep{Bedding2020,Murphy2020}.

\section{Conclusions}

We have examined 434 stars observed by the \tess\ mission, drawn from the catalogues of \dsct\ stars by \citet{Rodriguez2000} and \citet{Chang2013}.  We found that 62 are not \dsct\ pulsators, with most instead showing variability from binarity. 
For each of the 372 \dsct\ stars, we measured the frequency and amplitude of their strongest pulsation mode.  Using Gaia DR3 parallaxes, we placed the stars in the period--luminosity diagram (Fig.~\ref{fig:PL_revised}), in which most stars lie on a ridge that corresponds to pulsation in the fundamental radial mode.
We showed the value of the period--luminosity diagram in distinguishing $\delta$ Scuti stars from short-period RR~Lyrae stars.
We also confirmed the findings of \citet{Ziaali2019} that many stars fall on a second ridge that is a factor two shorter in period.  This second ridge is seen more clearly than before, thanks for the revised periods and distances.  Given that several stars falling on this ridge undergo amplitude modulation on timescales of years, this may signify energy transfer between modes and suggests a transient nature to a star's occupation of the second ridge.

Using densities published by \citet{Berger2020}, we show that the $\delta$ Scuti stars observed by \kepler\ follow a tight period--density relation, with a pulsation constant for the fundamental mode of $Q=0.0315$\,d.  Finally, we discuss several new examples of high-frequency $\delta$ Scuti stars with regular sequences of overtone modes, including XX Pyx and 29 Cyg.  

\section*{Acknowledgements}

We gratefully acknowledge support from the Australian Research Council through DECRA grant DE180101104, Future Fellowship FT210100485, and Discovery Project DP210103119, and
from the Danish National Research Foundation (Grant DNRF106) through its
funding for the Stellar Astrophysics Center (SAC).
This work has made use of data from the European Space Agency (ESA) mission {\em Gaia}, (\url{https://www.cosmos.esa.int/gaia}), 
processed by the {\em Gaia} Data Processing and Analysis Consortium (DPAC, \url{https://www.cosmos.esa.int/web/gaia/dpac/consortium}). Funding for the DPAC has been provided by national institutions, in particular the institutions participating in the {\em Gaia} Multilateral Agreement.
We are grateful to the entire Gaia and \kepler\ teams for providing the data used in this paper.
We also thank L\'aszl\'o Moln\'ar for useful discussions on RR~Lyrae stars.

\section*{Data Availability}

The \tess\ data underlying this article are available at the MAST Portal (Barbara A. Mikulski Archive for Space Telescopes), at \url{https://mast.stsci.edu/portal/Mashup/Clients/Mast/Portal.html}

% \bibliographystyle{mnras}
% \bibliography{references}
\input{output.bbl} % for arxiv version

%TABLE

\onecolumn

\begin{longtable}{rlrcrrrrrrrrl}
\caption{Properties of 372 \dsct\ stars in our sample. Column R2000 is set to 1 for stars listed in \citet{Rodriguez2000}, while the remainder are from \citet{Chang2013}; $f_{\rm cat}$ is the published dominant frequency; $f_1$ and $a_1$ are the frequency and amplitude of the dominant frequency we measured with \tess; the remaining columns were calculated as described in Sec.~\ref{sec:PL}.}
\label{tab:freq_table}\\ 
\hline
%HD & Name & TIC & FFI & Sectors & f\textsubscript{cat} & $f_{1}$ & $a_{1}$ & plx & dist & $V$ & $A_{v}$ & $M_{v}$ & e$\_M_{v}$\\
 %& & & & & \cd\ & \cd\ & ppt & mas & pc & mag & mag & mag & \\ 

HD & Name & TIC & R2000 &  f\textsubscript{cat} & $f_{1}$ & $a_{1}$ & dist & e\_dist & $A_{V}$ & e$\_A_{V}$ & $M_{V}$ & e$\_M_{V}$\\
 & & & & \cd\ & \cd\ & ppt & pc & pc & mag & mag & mag & mag\\
 
\hline
\endfirsthead

\multicolumn{13}{c}
{{\bfseries \tablename\ \thetable{} --- continued from previous page}} \\
\hline
%HD & Name & TIC & FFI & Sectors & f\textsubscript{cat} & $f_{1}$ & $a_{1}$ & plx & dist & $V$ & $A_{v}$ & $M_{v}$ & e$\_M_{v}$\\
% & & & & & \cd\ & \cd\ & ppt & mas & pc & mag & mag & mag & \\ 
 
HD & Name & TIC & R2000 &  f\textsubscript{cat} & $f_{1}$ & $a_{1}$ & dist & e\_dist & $A_{V}$ & e$\_A_{V}$ & $M_{V}$ & e$\_M_{V}$\\
 & & & & \cd\ & \cd\ & ppt & pc & pc & mag & mag & mag & mag\\
 \hline
\endhead

\hline
\hline
\endlastfoot

\input{final_star_sample.tex}

\end{longtable}

% \twocolumn

%%%%%%%%%%%%%%%%%%%%%%%%%%%%%%%%%%%%%%%%%%%%%%%%%%

% Don't change these lines
\bsp	% typesetting comment
\label{lastpage}
\end{document}

%% file: removed_star_sample.tex
\begin{tabular}{rlrc}
\toprule
    HD &      Name &    TIC &  R2000 \\
\midrule
  2145 &         BQ Phe &   7203984 &      1 \\
  4173 &       V756 Cas & 420842854 &      0 \\
 10088 &            --- & 151056397 &      1 \\
 10270 & CE Hyi         & 230983365 &      1 \\
 21102 &      V1241 Tau & 279181220 &      1 \\
 22544 &       V579 Per & 200937303 &      1 \\
 25637 &         AK Men & 394699148 &      1 \\
 43378 &         UZ Lyn & 322679290 &      0 \\
 43601 &            --- & 141770299 &      0 \\
 50420 &       V352 Aur & 191999705 &      1 \\
 60987 &       V345 Gem &   4164713 &      1 \\
 77140 &         FZ Vel &  30906332 &      1 \\
 77140 &         FZ Vel &  30906332 &      0 \\
 90747 &         GS UMa &   8245065 &      1 \\
100495 &      V1228 Cen & 318670307 &      1 \\
101065 &     V0816 Cen  & 163587609 &      0 \\
106103 &     GM Com     &   4738909 &      0 \\
108662 &     AI Com     & 393808105 &      0 \\
108945 &     UU Com     & 393819751 &      0 \\
109738 &            --- & 327436923 &      0 \\
110411 &     $\rho$ Vir & 390607705 &      0 \\
127986 &         CP Boo &  67993023 &      0 \\
129723 &         BP Oct & 290356320 &      0 \\
152896 &       V645 Her & 458574139 &      0 \\
165373 &       V831 Her & 135362677 &      0 \\
191747 &         18 Vul & 245152467 &      1 \\
191803 &         DE Oct & 261459125 &      1 \\
193084 &            --- & 164328483 &      1 \\
202444 &     $\tau$ Cyg & 167092249 &      0 \\
204420 &            --- & 301795154 &      0 \\
213272 &            --- & 100601011 &      1 \\
213272 &            --- & 100601011 &      0 \\
229085 &            --- &  13332837 &      0 \\
271923 &                &  41232835 &      0 \\
289243 &     V0783 Mon  & 237554493 &      0 \\
302013 &       V753 Cen & 300892362 &      1 \\
339669 & V382 Vul       & 245236103 &      1 \\
343341 & V1003 Her      & 282137421 &      1 \\
   --- &         TU UMi & 257002707 &      1 \\
   --- &         TV Lyn & 157798407 &      1 \\
   --- &         UY Cam & 441619744 &      1 \\
   --- &       V705 Per & 101256059 &      0 \\
   --- &            --- &  25671619 &      0 \\
   --- &            --- &  34137913 &      0 \\
   --- &            --- &  35166984 &      0 \\
   --- &            --- &  41994216 &      0 \\
   --- &            --- &  60657338 &      0 \\
   --- &            --- &  70870736 &      0 \\
   --- &            --- &  73764693 &      0 \\
   --- &            --- &  78002882 &      0 \\
   --- &            --- & 138735041 &      0 \\
   --- &            --- & 152288704 &      0 \\
   --- &            --- & 152359020 &      0 \\
   --- &            --- & 185259483 &      0 \\
   --- &            --- & 238170857 &      0 \\
   --- &            --- & 248348545 &      0 \\
   --- &            --- & 270709500 &      0 \\
   --- &            --- & 335826251 &      0 \\
   --- &            --- & 383604347 &      0 \\
   --- &            --- & 396424970 &      0 \\
   --- &            --- & 410299018 &      0 \\
   --- &            --- & 421452112 &      0 \\
\bottomrule
\end{tabular}

%% file: final_star_sample.tex
205 &       V1026 Cas & 359127550 &          0 &      --- &  6.25 & 0.01 &  307.96 &   1.16 & 0.10 &  0.06 &  1.21 &  0.06 \\
  1097 &          AU Scl &  12473170 &          1 &    17.73 & 16.55 & 0.00 &  325.34 &   5.44 & 0.13 &  0.05 &  1.37 &  0.07 \\
  1479 &  V377 Cas       & 327585336 &          1 &    33.33 & 15.46 & 0.00 &     --- &    --- & 0.00 &  0.00 &  2.25 &  0.00 \\
  2628 &          GN And & 440665786 &          1 &    14.43 & 14.43 & 0.01 &   61.23 &   0.54 & 0.05 &  0.04 &  1.24 &  0.04 \\
  2724 &          BB Phe & 116157537 &          1 &     5.74 &  6.16 & 0.00 &  134.51 &   0.38 & 0.11 &  0.05 &  0.42 &  0.05 \\
  3112 &         tet Tuc &  38847248 &          1 &    20.28 & 20.28 & 0.01 &  139.98 &   0.44 & 0.07 &  0.05 &  0.32 &  0.05 \\
  3326 &          BG Cet &  98660068 &          1 &    33.44 & 30.52 & 0.00 &   50.66 &   0.09 & 0.06 &  0.03 &  2.48 &  0.03 \\
  4490 &          XX Psc & 435860104 &          1 &     9.62 & 12.99 & 0.00 &  102.20 &   0.28 & 0.06 &  0.04 &  1.00 &  0.04 \\
  4494 &          CN Phe &  80432574 &          1 &    14.29 & 12.83 & 0.00 &  266.09 &   1.22 & 0.07 &  0.03 &  2.26 &  0.04 \\
  4849 &          AZ Phe &  80474886 &          1 &    18.15 & 18.17 & 0.00 &   97.06 &   0.63 & 0.20 &  0.07 &  1.35 &  0.07 \\
  4919 &      $\rho$ Phe & 369666165 &          1 &     5.45 & 10.75 & 0.01 &   75.00 &   0.29 & 0.13 &  0.10 &  0.73 &  0.11 \\
  6667 &             --- & 248345301 &          0 &    18.02 & 18.02 & 0.01 &  255.76 &   6.46 & 0.06 &  0.05 &  2.36 &  0.07 \\
  6859 &        V361 And & 186297798 &          1 &     8.76 &  8.77 & 0.01 &  238.84 &   1.45 & 0.11 &  0.07 &  0.74 &  0.07 \\
  6870 &          BS Tuc & 234548714 &          1 &    15.38 & 17.87 & 0.00 &  107.55 &   0.19 & 0.04 &  0.03 &  2.28 &  0.03 \\
  7312 &          AI Scl & 183595451 &          1 &    24.04 & 10.71 & 0.00 &   68.68 &   0.11 & 0.06 &  0.04 &  1.70 &  0.04 \\
  8511 &          AV Cet &  10838265 &          1 &    14.60 & 15.91 & 0.00 &   66.25 &   0.13 & 0.04 &  0.03 &  2.07 &  0.03 \\
  8781 &          BG Hyi &  52258534 &          1 &     8.96 &  9.04 & 0.00 &  186.10 &   0.47 & 0.11 &  0.05 &  1.61 &  0.05 \\
  8801 &        V529 And & 189381343 &          0 &    20.66 &  9.36 & 0.00 &   53.30 &   0.00 & 0.06 &  0.04 &  2.73 &  0.04 \\
  9065 &          WZ Scl &  70489756 &          1 &    10.43 & 10.42 & 0.01 &   93.51 &   0.21 & 0.12 &  0.04 &  1.61 &  0.04 \\
  9100 &          VX Psc & 381320713 &          1 &     7.35 &  6.34 & 0.01 &  138.79 &   0.61 & 0.05 &  0.04 &  0.26 &  0.04 \\
  9133 &          XX Scl &  70490868 &          1 &    20.45 & 20.36 & 0.00 &  334.22 &   6.03 & 0.02 &  0.02 &  1.27 &  0.04 \\
  9800 &        V365 And & 308719469 &          1 &     7.00 &  7.00 & 0.01 &  168.37 &   0.53 & 0.11 &  0.07 &  1.15 &  0.08 \\
 10845 &  VY Psc         &  88773937 &          1 &     4.57 &  5.51 & 0.00 &  154.27 &   0.40 & 0.05 &  0.04 &  0.55 &  0.05 \\
 11285 &          VV Ari &  91277922 &          1 &    13.09 & 12.63 & 0.00 &  168.27 &   1.07 & 0.10 &  0.05 &  0.48 &  0.05 \\
 11413 &          BD Phe & 229150702 &          1 &    26.81 & 26.80 & 0.01 &   78.04 &   0.17 & 0.03 &  0.03 &  1.44 &  0.03 \\
 11522 &          BK Cet &  92984630 &          1 &    11.07 & 11.06 & 0.00 &  101.36 &   0.32 & 0.07 &  0.04 &  0.69 &  0.04 \\
 11667 &             --- & 229154157 &          0 &    10.99 & 10.99 & 0.08 &  457.77 &   2.32 & 0.06 &  0.05 &  1.38 &  0.06 \\
 11956 &          FG Eri & 231048083 &          1 &     6.29 &  6.29 & 0.01 &  230.32 &   0.91 & 0.03 &  0.02 & -0.12 &  0.02 \\
 12389 &             --- &  63395663 &          1 &    25.00 & 20.83 & 0.01 &  229.70 &   2.23 & 0.03 &  0.03 &  1.14 &  0.04 \\
 12743 &        V373 And & 184589181 &          1 &     8.87 &  9.24 & 0.01 &  153.77 &   0.49 & 0.10 &  0.05 &  1.54 &  0.05 \\
 13079 &        V419 And & 184679514 &          1 &    20.00 & 19.41 & 0.00 &  181.74 &   0.64 & 0.08 &  0.06 &  2.53 &  0.06 \\
 13122 &        V784 Cas &  12221925 &          1 &     9.16 &  9.16 & 0.02 &   98.18 &   0.27 & 0.10 &  0.06 &  1.58 &  0.06 \\
 13755 &          CV Phe &   7245720 &          0 &    12.50 &  7.93 & 0.01 &  200.20 &   0.65 & 0.08 &  0.05 &  1.25 &  0.05 \\
 15634 &          TY For & 120932152 &          1 &    10.30 & 10.30 & 0.00 &   91.93 &   0.85 & 0.04 &  0.04 &  1.67 &  0.04 \\
 16189 &          DX Cet & 278962831 &          1 &     9.62 &  9.62 & 0.05 &  109.16 &   0.32 & 0.07 &  0.06 &  1.75 &  0.06 \\
 16439 &        V663 Cas & 280129361 &          1 &    17.12 & 17.13 & 0.01 &  309.04 &   1.52 & 0.03 &  0.02 &  1.21 &  0.03 \\
 16698 &          FI Eri & 142266756 &          1 &     5.84 &  5.84 & 0.03 &  324.29 &   3.10 & 0.08 &  0.05 &  0.83 &  0.06 \\
 16723 &          BS Cet & 441131892 &          1 &     8.93 &  7.61 & 0.01 &  147.60 &   0.50 & 0.07 &  0.05 &  0.79 &  0.05 \\
 17138 &          RZ Cas & 302771290 &          1 &    64.10 & 64.20 & 0.00 &   65.31 &   0.10 & 0.03 &  0.02 &  2.17 &  0.03 \\
 17892 &        V579 Per &  67712263 &          1 &    20.24 & 18.55 & 0.00 &     --- &    --- & 0.00 &  0.00 &  1.34 &  0.01 \\
 17978 &          VV For &  65484200 &          1 &    17.33 & 11.78 & 0.00 &  324.82 &   1.89 & 0.05 &  0.04 &  2.11 &  0.05 \\
 18655 &             --- & 365225836 &          0 &    14.88 & 14.88 & 0.02 &  283.54 &   1.74 & 0.07 &  0.05 &  2.33 &  0.06 \\
 18878 &        V509 Per & 192374290 &          1 &     6.86 &  6.87 & 0.01 &  115.94 &   0.35 & 0.07 &  0.04 &  1.08 &  0.05 \\
 19279 &        V521 Per & 192533869 &          1 &    14.45 & 14.85 & 0.00 &   86.13 &   0.25 & 0.02 &  0.02 &  1.68 &  0.02 \\
 20313 &          BN Hyi & 348762920 &          1 &    15.15 & 12.31 & 0.00 &   80.27 &   0.28 & 0.07 &  0.05 &  1.08 &  0.05 \\
 20429 &          AR Ari &  28833184 &          1 &     5.65 &  5.15 & 0.02 &  239.09 &   2.01 & 0.11 &  0.09 &  0.81 &  0.09 \\
 20919 &        V461 Per & 252829836 &          1 &    28.57 & 28.04 & 0.00 &  181.26 &   1.07 & 0.09 &  0.04 &  2.64 &  0.05 \\
 21190 &          CP Oct & 348772511 &          1 &     6.68 &  6.68 & 0.01 &  200.14 &   2.48 & 0.13 &  0.07 &  1.00 &  0.08 \\
 21553 &        V465 Per & 347570557 &          1 &    14.04 & 14.04 & 0.00 &  180.75 &   0.61 & 0.04 &  0.03 &  2.44 &  0.04 \\
 21985 &          AS Eri & 301407485 &          1 &    58.82 & 59.03 & 0.00 &  210.64 &   1.44 & 0.03 &  0.03 &  1.66 &  0.03 \\
 22541 &          AD Hor &  79611815 &          1 &    10.99 & 10.99 & 0.02 &  203.79 &   0.76 & 0.03 &  0.02 &  0.45 &  0.03 \\
 23728 &        V376 Per & 432049507 &          1 &    10.06 & 10.06 & 0.02 &   67.90 &   0.77 & 0.08 &  0.07 &  1.73 &  0.07 \\
 24550 &        V479 Tau & 459908110 &          1 &    13.19 & 13.23 & 0.00 &  152.57 &   0.44 & 0.17 &  0.06 &  1.34 &  0.06 \\
 24809 &        V386 Per &  94549636 &          1 &    18.18 & 22.33 & 0.00 &   63.84 &   0.09 & 0.01 &  0.01 &  2.50 &  0.02 \\
 26123 &  V407 Cep       & 420755646 &          1 &    13.57 & 13.58 & 0.02 &     --- &    --- & 0.00 &  0.00 &  1.96 &  0.01 \\
 26574 &          38 Eri &  67687505 &          1 &    13.39 & 12.23 & 0.01 &   38.42 &   0.22 & 0.22 &  0.17 &  0.90 &  0.17 \\
 26892 &          UZ Ret &  38515566 &          1 &     8.14 &  8.14 & 0.01 &  339.81 &   1.18 & 0.12 &  0.06 &  1.45 &  0.07 \\
 27093 &          IU Eri &  67781014 &          0 &    14.51 & 12.48 & 0.00 &  226.69 &   1.08 & 0.03 &  0.02 &  0.68 &  0.03 \\
 27397 &        V483 Tau & 435910664 &          1 &    18.21 & 17.25 & 0.00 &   45.48 &   0.11 & 0.06 &  0.05 &  2.23 &  0.05 \\
 27459 &        V696 Tau & 435916016 &          1 &    27.32 & 24.77 & 0.00 &   47.40 &   0.24 & 0.03 &  0.03 &  1.84 &  0.03 \\
 27503 &          BR Hyi &  25195864 &          1 &     5.00 & 26.74 & 0.00 &  664.38 &   4.97 & 0.06 &  0.05 &  1.61 &  0.09 \\
 27545 &          TX Ret &  38587180 &          1 &    14.93 &  9.63 & 0.00 &  199.33 &   0.54 & 0.07 &  0.04 &  1.40 &  0.04 \\
 27628 &        V775 Tau & 435923755 &          1 &    15.87 & 13.08 & 0.00 &   45.42 &   0.14 & 0.09 &  0.05 &  2.34 &  0.05 \\
 28052 &        V777 Tau &  60879864 &          1 &     5.49 &  5.18 & 0.00 &   46.72 &   0.48 & 0.05 &  0.04 &  1.08 &  0.05 \\
 28665 &          EQ Eri & 178880326 &          1 &    14.29 & 14.37 & 0.02 &  212.39 &   0.83 & 0.08 &  0.05 &  1.01 &  0.05 \\
 28837 &          RX Cae &   7808834 &          1 &     6.49 &  6.49 & 0.03 &  147.51 &   1.36 & 0.13 &  0.08 &  1.04 &  0.08 \\
 28910 &      $\rho$ Tau & 245860427 &          1 &    14.93 & 15.97 & 0.00 &   46.31 &   0.29 & 0.03 &  0.03 &  1.29 &  0.03 \\
 29870 &      HS Eri     & 298981924 &          0 &      --- &  6.89 & 0.01 &  124.29 &   0.37 & 0.12 &  0.08 &  1.15 &  0.09 \\
 30422 &          EX Eri &    589826 &          1 &    47.62 & 48.09 & 0.00 &   57.13 &   0.06 & 0.04 &  0.03 &  2.35 &  0.03 \\
 30600 &          HV Eri & 167486272 &          1 &     4.74 &  4.74 & 0.02 &  339.75 &   2.59 & 0.06 &  0.05 &  0.63 &  0.06 \\
 30716 &       V1359 Ori &  11083205 &          1 &     5.49 &  5.49 & 0.01 &  323.51 &   3.90 & 0.10 &  0.06 &  0.82 &  0.07 \\
 31908 &  XZ Men         & 319289587 &          1 &     9.23 & 11.00 & 0.01 &     --- &    --- & 0.00 &  0.00 &  0.30 &  0.13 \\
 32045 &           S Eri & 152373997 &          1 &     3.66 &  3.27 & 0.00 &   86.75 &   0.60 & 0.06 &  0.05 &  0.03 &  0.06 \\
 32846 &           X Cae &  77669416 &          1 &     7.40 &  7.39 & 0.02 &  104.70 &   0.57 & 0.05 &  0.05 &  1.17 &  0.05 \\
 33959 &          KW Aur &   1840666 &          1 &    11.35 & 11.35 & 0.01 &   83.24 &   0.72 & 0.05 &  0.05 &  0.36 &  0.05 \\
 34409 &          BS Cam & 417652271 &          1 &     5.54 &  4.17 & 0.00 &  401.04 &   2.87 & 0.11 &  0.06 &  0.31 &  0.06 \\
 37857 &          EE Cam &  70657495 &          1 &     4.93 &  4.93 & 0.02 &  223.94 &   0.88 & 0.11 &  0.09 &  0.84 &  0.09 \\
 38882 &          RY Lep &  93441696 &          1 &     4.44 &  4.44 & 0.10 &  437.06 &  15.66 & 0.13 &  0.11 & -0.06 &  0.13 \\
 39244 &          YY Pic & 299884840 &          1 &     9.74 &  9.74 & 0.02 &  153.22 &   0.36 & 0.04 &  0.03 &  1.57 &  0.03 \\
 39996 &          AA Col & 100830119 &          1 &     6.66 &  6.66 & 0.02 &  226.17 &   0.80 & 0.11 &  0.07 &  1.27 &  0.07 \\
 40372 &       V1004 Ori & 282270717 &          1 &    16.39 & 15.31 & 0.00 &  106.63 &   0.37 & 0.04 &  0.03 &  0.71 &  0.03 \\
 40535 &        V474 Mon &  67265166 &          1 &     7.35 &  7.35 & 0.06 &   93.82 &   0.26 & 0.09 &  0.07 &  1.19 &  0.07 \\
 40765 &          UY Col & 143168754 &          1 &     5.90 &  5.90 & 0.04 &  474.20 &   2.46 & 0.12 &  0.09 &  1.05 &  0.09 \\
 42503 &      AU Col     & 300114686 &          0 &     7.00 &  5.25 & 0.00 &  237.38 &   0.21 & 0.05 &  0.04 &  0.52 &  0.04 \\
 44958 &        V435 Car & 255548143 &          1 &    20.00 & 20.57 & 0.00 &   71.29 &   0.08 & 0.03 &  0.03 &  2.44 &  0.03 \\
 45311 &        V456 Aur & 144101735 &          1 &     7.21 &  7.21 & 0.02 &  187.98 &   1.41 & 0.11 &  0.07 &  1.36 &  0.07 \\
 50018 &          OX Aur &  21291018 &          1 &     6.46 &  6.80 & 0.00 &  147.78 &   0.93 & 0.15 &  0.07 &  0.11 &  0.08 \\
 52788 &        V383 Car & 279361762 &          1 &     8.33 &  8.91 & 0.00 &  303.99 &   1.08 & 0.14 &  0.07 &  0.86 &  0.07 \\
 54250 &       V752 Mon  & 168524254 &          0 &     4.32 &  5.32 & 0.01 &     --- &    --- & 0.00 &  0.00 &  0.95 &  0.01 \\
 55057 &        V571 Mon &   5456605 &          1 &    10.00 &  5.25 & 0.01 &   87.49 &   0.49 & 0.08 &  0.05 &  0.65 &  0.06 \\
 55595 &          HN CMa &  65472324 &          1 &     4.00 &  4.73 & 0.00 &  211.84 &   0.86 & 0.04 &  0.02 & -0.07 &  0.03 \\
 57167 &  R CMa          & 409319605 &          1 &    21.28 & 15.85 & 0.00 &     --- &    --- & 0.00 &  0.00 &  2.55 &  0.01 \\
 58634 &        V368 Pup & 173139593 &          1 &     8.67 &  8.67 & 0.01 &  180.87 &   0.57 & 0.02 &  0.02 &  0.56 &  0.02 \\
 58954 &          NR CMa &  49832986 &          1 &     6.02 &  6.85 & 0.01 &   89.97 &   0.27 & 0.09 &  0.05 &  0.77 &  0.05 \\
 59594 &        V349 Pup & 112484997 &          1 &    19.88 & 19.87 & 0.01 &   93.34 &   0.16 & 0.02 &  0.02 &  2.45 &  0.02 \\
 60302 &        V344 Gem & 247165103 &          1 &     8.46 &  8.46 & 0.01 &  185.60 &   0.81 & 0.07 &  0.04 &  1.62 &  0.04 \\
 62437 &          AZ CMi & 280680714 &          1 &    10.49 & 10.50 & 0.02 &  148.81 &   1.09 & 0.05 &  0.03 &  0.55 &  0.04 \\
 63148 &       V391 Pup  & 126928121 &          0 &      --- &  8.73 & 0.01 &     --- &    --- & 0.00 &  0.00 &  0.49 &  0.01 \\
 63874 &          ET Lyn &  39011225 &          0 &     5.42 &  5.42 & 0.00 &  213.93 &   8.51 & 0.09 &  0.05 &  0.66 &  0.08 \\
 64191 &          AD CMi & 266328148 &          1 &     8.13 &  8.13 & 0.09 &  451.67 &   2.90 & 0.08 &  0.06 &  0.97 &  0.07 \\
 64491 &          DD Lyn &  14878438 &          1 &    20.41 & 19.76 & 0.00 &   64.93 &   0.32 & 0.08 &  0.04 &  2.08 &  0.04 \\
 66260 &        V393 Car & 364399376 &          1 &     7.08 &  7.08 & 0.06 &  199.66 &   3.54 & 0.07 &  0.05 &  0.90 &  0.06 \\
 67028 &             --- & 452926063 &          1 &    19.31 & 19.32 & 0.00 &  431.76 &   4.37 & 0.05 &  0.03 &  0.59 &  0.04 \\
 67147 &          CO Lyn &  80951366 &          1 &     6.11 & 12.67 & 0.01 &  141.61 &   0.39 & 0.10 &  0.06 &  0.97 &  0.06 \\
 67290 &        V355 Pup & 176152984 &          1 &     6.71 &  7.10 & 0.01 &  368.17 &   2.65 & 0.05 &  0.03 &  0.24 &  0.04 \\
 67390 &          SZ Lyn & 192939152 &          1 &     8.30 &  8.30 & 0.14 &  525.06 &  25.47 & 0.10 &  0.09 &  0.74 &  0.13 \\
 67523 &      $\rho$ Pup & 154360594 &          1 &     7.10 &  7.10 & 0.02 &   19.44 &   0.08 & 0.13 &  0.12 &  1.22 &  0.12 \\
 67852 &      DE CMi     & 452982723 &          0 &    11.30 & 11.30 & 0.01 &     --- &    --- & 0.00 &  0.00 &  1.86 &  0.01 \\
 67911 &          CQ Lyn &  81003138 &          1 &     8.87 &  8.87 & 0.02 &  196.57 &   1.57 & 0.10 &  0.07 &  1.40 &  0.07 \\
 69213 &          AI Vel &  81709032 &          1 &     8.96 &  8.96 & 0.09 &  100.30 &   0.17 & 0.05 &  0.05 &  1.52 &  0.05 \\
 69242 &          CR Lyn &  22684612 &          1 &     7.59 &  7.59 & 0.01 &  172.91 &   0.65 & 0.10 &  0.07 &  1.36 &  0.07 \\
 69997 &          HQ Hya &  19566581 &          1 &    13.25 & 11.71 & 0.00 &  110.32 &   0.24 & 0.12 &  0.05 &  0.98 &  0.05 \\
 71297 &          LM Hya & 169051549 &          1 &    26.32 & 26.30 & 0.00 &   46.74 &   1.25 & 0.09 &  0.14 &  2.16 &  0.15 \\
 71935 &          GU Vel & 140349824 &          1 &    14.29 & 15.01 & 0.00 &   60.66 &   0.27 & 0.04 &  0.04 &  1.12 &  0.04 \\
 73857 &          VZ Cnc & 366632312 &          1 &     5.61 &  5.61 & 0.12 &  221.97 &   1.15 & 0.11 &  0.07 &  0.86 &  0.08 \\
 74292 &          FL Cnc & 165859518 &          1 &    12.50 & 13.41 & 0.00 &   97.47 &   0.25 & 0.05 &  0.03 &  2.04 &  0.03 \\
 75654 &          HZ Vel & 190158777 &          1 &    11.49 & 14.34 & 0.00 &   71.85 &   0.09 & 0.04 &  0.03 &  2.05 &  0.03 \\
 75747 &          RS Cha & 323292655 &          1 &    12.50 & 11.62 & 0.00 &   98.63 &   0.18 & 0.06 &  0.06 &  1.06 &  0.06 \\
 77906 &             --- & 401744720 &          0 &    12.90 & 12.91 & 0.07 &  344.85 &   7.24 & 0.09 &  0.06 &  1.99 &  0.08 \\
 78422 &          NT Hya &  20042408 &          1 &     6.94 &  6.90 & 0.01 &  165.86 &   0.59 & 0.09 &  0.06 &  1.19 &  0.06 \\
 79185 &          MP Vel &  74528796 &          1 &     4.29 &  7.92 & 0.00 &  269.47 &   1.13 & 0.10 &  0.05 &  0.60 &  0.06 \\
 79439 &          DD UMa & 137648019 &          1 &     8.00 & 25.37 & 0.00 &   35.83 &   0.10 & 0.03 &  0.02 &  2.01 &  0.03 \\
 79781 &          GG UMa &  86232782 &          1 &     7.42 &  7.42 & 0.02 &  266.90 &   1.00 & 0.12 &  0.07 &  1.35 &  0.07 \\
 79889 &          BE Lyn &  56914404 &          1 &    10.43 & 10.43 & 0.11 &  251.63 &   1.31 & 0.04 &  0.04 &  1.76 &  0.05 \\
 80886 &             --- & 452464329 &          0 &    14.27 & 14.26 & 0.02 &     --- &    --- & 0.00 &  0.00 &  1.83 &  0.01 \\
 81882 &          KZ UMa & 147893556 &          1 &    16.67 & 14.28 & 0.00 &  229.55 &   0.97 & 0.06 &  0.04 &  1.26 &  0.04 \\
 82620 &          DL UMa & 147574768 &          1 &    12.03 & 13.08 & 0.01 &  126.72 &   0.26 & 0.10 &  0.04 &  1.57 &  0.04 \\
 83041 &          AK Ant & 189569123 &          1 &    15.15 & 14.53 & 0.00 &  305.90 &   2.22 & 0.08 &  0.04 &  1.29 &  0.05 \\
 84712 &          MT Vel &  34156938 &          1 &    12.50 & 13.60 & 0.01 &  240.74 &   1.50 & 0.08 &  0.04 &  0.62 &  0.05 \\
 84800 &          IX UMa &  23969903 &          1 &    41.67 & 36.62 & 0.00 &  147.98 &   0.43 & 0.05 &  0.03 &  1.88 &  0.04 \\
 84948 &             --- & 453432481 &          1 &    12.82 & 12.63 & 0.00 &  255.19 &   3.95 & 0.06 &  0.04 &  1.03 &  0.05 \\
 84999 &         $\upsilon$ UMa & 331900366 &          1 &     7.54 &  6.27 & 0.01 &   35.67 &   0.24 & 0.09 &  0.08 &  0.92 &  0.08 \\
 85040 &          DG Leo &  88024537 &          1 &    12.11 & 12.11 & 0.00 &  216.27 &  22.95 & 0.03 &  0.03 & -0.36 &  0.18 \\
 86301 &          BF Ant &  22234795 &          0 &      --- &  5.03 & 0.00 &  143.91 &   0.47 & 0.03 &  0.02 &  0.49 &  0.03 \\
 87700 &        V336 Vel & 462426434 &          1 &     8.27 & 12.42 & 0.01 &  132.24 &   0.26 & 0.13 &  0.06 &  1.11 &  0.06 \\
 88824 &          LW Vel & 259761784 &          1 &     7.98 &  8.98 & 0.01 &   49.67 &   0.13 & 0.05 &  0.04 &  1.74 &  0.04 \\
 89343 &          EN UMa & 103692670 &          1 &     9.09 &  6.43 & 0.01 &  130.07 &   0.69 & 0.06 &  0.04 &  0.24 &  0.05 \\
 90001 &        V344 Vel & 220628281 &          1 &     6.68 &  6.68 & 0.01 &  247.53 &   1.20 & 0.08 &  0.05 &  0.84 &  0.05 \\
 93044 &          EO UMa &  17329905 &          1 &    11.90 & 11.91 & 0.02 &  113.97 &   0.28 & 0.06 &  0.04 &  1.78 &  0.04 \\
 93137 &          UX LMi & 337319844 &          1 &     6.64 &  6.64 & 0.02 &  426.41 &  10.83 & 0.11 &  0.08 &  0.85 &  0.09 \\
 93142 &          AZ Ant &  54253999 &          1 &     9.50 &  9.50 & 0.01 &  256.99 &   1.41 & 0.10 &  0.05 &  0.73 &  0.06 \\
 93298 &        V353 Vel & 106886169 &          1 &     4.04 &  4.04 & 0.01 &  312.77 &   1.88 & 0.04 &  0.03 &  0.12 &  0.04 \\
 94033 &          KZ Hya & 188209486 &          1 &    16.81 & 16.80 & 0.22 &  330.80 &   7.19 & 0.11 &  0.09 &  2.36 &  0.11 \\
 94480 &          WW LMi & 138738145 &          0 &      --- &  7.88 & 0.01 &  130.68 &   0.55 & 0.10 &  0.05 &  0.51 &  0.05 \\
 94985 &          IW Vel & 120856811 &          1 &     6.67 & 10.15 & 0.01 &  108.68 &   0.26 & 0.03 &  0.02 &  0.69 &  0.03 \\
 95321 &        V527 Car & 304050480 &          1 &     4.68 &  4.68 & 0.02 &  531.78 &   6.11 & 0.05 &  0.05 &  0.42 &  0.06 \\
 97302 &          FI UMa & 302349701 &          1 &    25.00 & 15.95 & 0.00 &  104.91 &   0.22 & 0.01 &  0.01 &  1.51 &  0.01 \\
 98851 &          LR UMa & 144311964 &          0 &    18.15 &  8.97 & 0.00 &  152.44 &   0.54 & 0.12 &  0.06 &  1.38 &  0.06 \\
 99002 &          CX UMa &  17992601 &          1 &    10.00 & 14.80 & 0.00 &  148.40 &   0.54 & 0.07 &  0.05 &  1.01 &  0.05 \\
 99983 &          HQ UMa & 284596378 &          1 &     8.65 &  8.65 & 0.01 &  163.82 &   1.36 & 0.11 &  0.05 &  0.93 &  0.05 \\
100363 &          SU Crt & 157787615 &          1 &    18.18 & 18.77 & 0.00 &  169.76 &   1.65 & 0.05 &  0.04 &  2.42 &  0.05 \\
101158 &        V837 Cen & 429143728 &          1 &    12.17 & 10.84 & 0.01 &  191.91 &   0.59 & 0.06 &  0.04 &  0.69 &  0.04 \\
101696 &          VY Crt & 270265624 &          1 &     7.34 &  7.34 & 0.03 &  131.88 &   2.89 & 0.12 &  0.08 &  1.17 &  0.09 \\
102355 &          KW UMa & 138590902 &          1 &     8.20 & 15.91 & 0.00 &  109.20 &   0.61 & 0.12 &  0.05 &  1.27 &  0.05 \\
102480 &      MX UMa     &  11891826 &          0 &     8.62 & 16.19 & 0.01 &  263.85 &   0.53 & 0.11 &  0.05 &  1.24 &  0.05 \\
102541 &       V1023 Cen & 454109915 &          1 &    20.00 & 19.90 & 0.01 &  115.78 &   0.31 & 0.03 &  0.02 &  2.60 &  0.03 \\
103279 &             --- & 293110952 &          0 &    14.20 & 14.21 & 0.03 &  421.82 &   8.48 & 0.06 &  0.05 &  2.71 &  0.10 \\
104036 &          EE Cha & 454961439 &          1 &    33.33 & 33.87 & 0.01 &  104.82 &   0.26 & 0.01 &  0.01 &  1.63 &  0.02 \\
104237 &          DX Cha & 357040880 &          1 &    33.33 & 33.28 & 0.00 &  106.52 &   0.43 & 0.03 &  0.03 &  1.43 &  0.04 \\
104288 &          OR Dra & 142946146 &          0 &     8.87 &  8.86 & 0.00 &  206.93 &  17.00 & 0.10 &  0.06 &  0.73 &  0.20 \\
104513 &  DP UMa         & 141201605 &          1 &    25.00 & 18.07 & 0.00 &     --- &    --- & 0.00 &  0.00 &  2.56 &  0.04 \\
105058 &             --- &  53523744 &          1 &    25.00 &  9.13 & 0.00 &     --- &    --- & 0.00 &  0.00 &  1.28 &  0.01 \\
105234 &          EF Cha & 357165560 &          1 &    18.87 & 17.95 & 0.00 &     --- &    --- & 0.00 &  0.00 &  2.41 &  0.01 \\
105513 &          CO Cru &  67533256 &          1 &     6.85 &  6.86 & 0.01 &  380.94 &   1.56 & 0.10 &  0.06 &  1.25 &  0.06 \\
105759 &          II Vir & 176867792 &          1 &    23.64 & 23.65 & 0.01 &   92.69 &   0.27 & 0.03 &  0.02 &  1.69 &  0.03 \\
107131 &          FM Com & 328860893 &          1 &    15.08 & 14.93 & 0.00 &   85.52 &   0.20 & 0.01 &  0.01 &  1.77 &  0.02 \\
107513 &          KU Com & 356702180 &          1 &    33.33 & 19.91 & 0.00 &   86.51 &   0.30 & 0.04 &  0.04 &  2.64 &  0.04 \\
107904 &          AI CVn & 284593009 &          1 &     8.60 &  5.05 & 0.02 &  131.81 &   1.32 & 0.10 &  0.07 &  0.33 &  0.07 \\
109585 &          TU Crv &  83207850 &          1 &    12.20 & 10.62 & 0.00 &   75.16 &   1.60 & 0.14 &  0.05 &  1.67 &  0.06 \\
109663 &             --- & 260842067 &          0 &     5.36 &  5.36 & 0.05 &  548.38 &   4.30 & 0.05 &  0.04 &  1.28 &  0.05 \\
110377 &          GG Vir & 390607729 &          1 &    20.00 & 31.35 & 0.00 &   71.65 &   0.16 & 0.01 &  0.01 &  1.93 &  0.02 \\
110951 &          FM Vir & 390633441 &          1 &    13.91 & 13.91 & 0.00 &   80.45 &   0.70 & 0.08 &  0.06 &  0.61 &  0.06 \\
111604 &      DT CVn     &  17654683 &          0 &     8.77 &  5.33 & 0.01 &  108.70 &   0.05 & 0.03 &  0.02 &  0.66 &  0.02 \\
111786 &          MO Hya &   9524429 &          1 &    31.06 & 36.01 & 0.00 &   62.09 &   0.43 & 0.04 &  0.03 &  2.14 &  0.04 \\
113537 &        V947 Cen & 453750774 &          1 &     7.29 &  7.30 & 0.01 &   99.95 &   0.28 & 0.12 &  0.07 &  1.31 &  0.07 \\
114042 &        V950 Cen & 253513495 &          1 &     6.61 &  6.61 & 0.00 &  258.12 &   1.66 & 0.04 &  0.02 &  0.67 &  0.03 \\
114620 &        V954 Cen & 442837381 &          1 &     9.14 &  9.14 & 0.02 &  139.80 &   0.31 & 0.09 &  0.07 &  1.44 &  0.07 \\
115308 &          DK Vir &  66693422 &          1 &     8.61 &  7.58 & 0.01 &  125.16 &   0.38 & 0.10 &  0.05 &  1.10 &  0.05 \\
115520 &             --- & 105579901 &          0 &    18.83 & 17.77 & 0.01 &  292.72 &   1.16 & 0.04 &  0.03 &  1.06 &  0.04 \\
115604 &          AO CVn & 180651551 &          1 &     8.22 &  8.22 & 0.01 &   78.25 &   0.64 & 0.07 &  0.07 &  0.17 &  0.07 \\
116994 &        V743 Cen & 438816206 &          1 &     9.78 &  9.78 & 0.05 &  226.64 &   1.08 & 0.07 &  0.06 &  1.85 &  0.06 \\
117589 &          PP Com & 368127841 &          0 &    16.05 & 17.72 & 0.00 &  106.41 &   0.03 & 0.07 &  0.03 &  2.17 &  0.03 \\
117661 &          HX Vir &  32146574 &          1 &    23.26 & 22.27 & 0.00 &   83.08 &   0.22 & 0.04 &  0.03 &  1.36 &  0.04 \\
118326 &          LW Mus & 342500794 &          1 &     6.51 &  6.51 & 0.04 &  375.31 &   8.82 & 0.14 &  0.09 &  1.21 &  0.10 \\
118743 &          BZ Boo & 282811788 &          0 &     2.70 & 16.30 & 0.00 &  225.58 &   1.73 & 0.10 &  0.04 &  1.31 &  0.04 \\
118954 &          IP UMa & 458452969 &          1 &    10.00 & 10.00 & 0.01 &  240.65 &   0.97 & 0.08 &  0.04 &  0.69 &  0.05 \\
120500 &          FQ Boo & 399182167 &          1 &    20.41 & 21.05 & 0.00 &  120.91 &   0.64 & 0.01 &  0.01 &  1.19 &  0.02 \\
120635 &             --- & 299948201 &          0 &     7.08 &  7.08 & 0.03 &  392.93 &   0.87 & 0.06 &  0.06 &  0.77 &  0.06 \\
120896 &          QT Vir & 392979958 &          0 &    17.79 &  7.61 & 0.00 &  223.10 &   0.98 & 0.07 &  0.04 &  1.70 &  0.04 \\
121517 &             --- & 241787384 &          0 &     9.01 & 11.01 & 0.07 &  300.62 &   1.26 & 0.09 &  0.08 &  1.93 &  0.08 \\
123460 &       V1338 Cen & 242302902 &          0 &     7.64 &  7.64 & 0.12 &  644.85 &   8.39 & 0.10 &  0.10 &  1.24 &  0.11 \\
124675 &       kap 2 Boo & 310362802 &          1 &    15.43 & 15.44 & 0.00 &   49.63 &   0.21 & 0.04 &  0.03 &  1.00 &  0.03 \\
124953 &          CN Boo & 135169898 &          1 &    22.83 & 11.56 & 0.00 &   46.20 &   0.06 & 0.04 &  0.03 &  2.60 &  0.03 \\
125081 &          MX Vir &  46041110 &          1 &     6.49 &  6.49 & 0.01 &  149.23 &   0.44 & 0.15 &  0.10 &  1.33 &  0.10 \\
125161 &     $\iota$ Boo & 310381204 &          1 &    37.74 & 36.75 & 0.00 &   29.51 &   0.06 & 0.03 &  0.03 &  2.36 &  0.03 \\
125162 &      lambda Boo & 168708816 &          1 &    43.48 & 38.17 & 0.00 &   30.64 &   0.12 & 0.03 &  0.03 &  1.71 &  0.03 \\
126742 &             --- & 241843363 &          0 &     5.99 &  5.99 & 0.05 &  550.00 &   3.50 & 0.16 &  0.11 &  0.89 &  0.12 \\
126859 &        V853 Cen & 290932786 &          1 &    18.94 & 16.31 & 0.00 &  140.57 &   0.25 & 0.15 &  0.04 &  1.09 &  0.04 \\
127269 &          MP Hya &  83860356 &          1 &    29.94 & 29.65 & 0.00 &  180.57 &   0.97 & 0.03 &  0.02 &  1.47 &  0.03 \\
127411 &          IT Dra & 166177270 &          1 &    23.04 & 16.85 & 0.00 &  139.30 &   0.32 & 0.04 &  0.03 &  1.76 &  0.04 \\
127695 &       V1034 Cen & 395108754 &          1 &     4.26 &  4.18 & 0.00 &  282.55 &   1.01 & 0.13 &  0.06 &  1.29 &  0.06 \\
127711 &       V1035 Cen & 395425079 &          1 &    12.50 & 11.68 & 0.00 &  276.62 &   2.37 & 0.16 &  0.04 &  1.34 &  0.05 \\
127759 &          EI Dra & 166179291 &          1 &    14.79 & 14.79 & 0.01 &  365.20 &   1.82 & 0.05 &  0.03 &  0.68 &  0.04 \\
127762 &         gam Boo &  67991192 &          1 &    13.70 & 13.59 & 0.00 &   26.29 &   0.12 & 0.07 &  0.05 &  0.85 &  0.05 \\
127927 &       V1036 Cen & 396855479 &          1 &     9.09 &  9.47 & 0.00 &  486.62 &   3.05 & 0.12 &  0.05 &  1.06 &  0.06 \\
127929 &          ER Dra & 166179517 &          1 &    11.35 & 11.38 & 0.00 &  122.26 &   0.33 & 0.07 &  0.03 &  0.77 &  0.04 \\
128157 &        V896 Cen & 430665759 &          1 &    20.00 & 19.86 & 0.00 &  152.75 &   0.46 & 0.11 &  0.05 &  2.26 &  0.05 \\
128862 &          PR Aps & 402305954 &          0 &      --- & 10.42 & 0.01 &  251.40 &   1.05 & 0.13 &  0.06 &  0.91 &  0.07 \\
129041 &          BT Cir & 403779926 &          1 &    28.25 & 28.20 & 0.00 &  114.10 &   0.27 & 0.03 &  0.02 &  2.10 &  0.02 \\
129494 &          BO Cir & 292435853 &          1 &     7.08 &  8.08 & 0.00 &  364.32 &   1.63 & 0.18 &  0.08 &  1.73 &  0.08 \\
132209 &          BV Cir & 453848870 &          1 &     6.33 &  6.33 & 0.01 &  127.77 &   0.30 & 0.10 &  0.05 &  0.94 &  0.05 \\
133194 &          HY Lib & 440958702 &          1 &     6.82 &  6.82 & 0.03 &  194.10 &   1.01 & 0.13 &  0.09 &  1.15 &  0.09 \\
134185 &          HU Lup & 122932611 &          1 &     6.52 &  7.66 & 0.01 &  454.35 &   2.64 & 0.11 &  0.07 &  0.77 &  0.07 \\
135383 &          OP Aps & 407012535 &          1 &     7.20 &  7.20 & 0.02 &  207.83 &   0.71 & 0.09 &  0.07 &  1.41 &  0.07 \\
140436 &      gam CrB    & 255861229 &          0 &    33.33 &  4.50 & 0.00 &   44.80 &   0.04 & 0.06 &  0.09 &  0.53 &  0.10 \\
140639 &             --- & 178463456 &          0 &     4.95 &  4.95 & 0.10 &  425.87 &   2.98 & 0.15 &  0.12 &  1.20 &  0.12 \\
143466 &          CL Dra & 458485558 &          1 &    13.11 & 14.85 & 0.00 &   33.78 &   0.08 & 0.05 &  0.04 &  2.27 &  0.05 \\
145393 &             --- & 382538797 &          1 &    31.95 & 33.61 & 0.00 &  251.10 &   0.33 & 0.11 &  0.06 &  2.76 &  0.06 \\
148638 &          NP TrA & 402646773 &          0 &    16.31 & 15.30 & 0.00 &  251.30 &   1.35 & 0.03 &  0.02 &  0.87 &  0.03 \\
149530 &       V1060 Sco & 280816457 &          1 &     5.53 &  5.58 & 0.01 &  404.33 &   2.83 & 0.15 &  0.09 &  0.64 &  0.09 \\
153747 &        V922 Sco & 346505765 &          1 &    21.28 & 21.47 & 0.01 &  177.14 &   0.69 & 0.04 &  0.03 &  1.14 &  0.03 \\
153805 &       V1072 Sco & 347028423 &          1 &     7.47 &  7.47 & 0.01 &  182.73 &   1.09 & 0.09 &  0.05 &  0.97 &  0.06 \\
154225 &  V929 Her       &  82551645 &          1 &     6.93 &  6.93 & 0.01 &     --- &    --- & 0.00 &  0.00 &  1.39 &  0.01 \\
154605 &             --- &  43173526 &          0 &     9.93 & 11.94 & 0.07 &  285.17 &   1.18 & 0.19 &  0.12 &  2.13 &  0.12 \\
154731 &      MR Dra     & 274674793 &          0 &    31.25 & 30.26 & 0.00 &  207.20 &   0.23 & 0.03 &  0.02 &  1.59 &  0.02 \\
155118 &        V873 Her & 143166570 &          1 &     7.91 &  7.86 & 0.00 &  240.52 &   1.61 & 0.10 &  0.07 &  1.40 &  0.07 \\
155514 &        V620 Her & 257456864 &          1 &    11.31 & 11.32 & 0.01 &   80.12 &   0.11 & 0.04 &  0.04 &  1.63 &  0.04 \\
158741 &        V949 Sco & 100701385 &          1 &     4.61 &  8.23 & 0.00 &  119.66 &   0.36 & 0.14 &  0.07 &  0.64 &  0.08 \\
159223 &        V648 Her & 321813050 &          0 &     3.45 & 46.30 & 0.00 &   79.24 &   0.10 & 0.03 &  0.03 &  2.32 &  0.04 \\
160589 &  V703 Sco       & 193310172 &          1 &     8.68 &  8.68 & 0.07 &  189.09 &   0.15 & 0.10 &  0.09 &  1.37 &  0.09 \\
161032 &        V352 Pav & 306889170 &          1 &     9.35 & 11.82 & 0.02 &  110.58 &   0.29 & 0.08 &  0.05 &  1.21 &  0.05 \\
161287 &        V966 Her & 256007029 &          1 &     7.52 &  7.52 & 0.00 &  212.82 &   0.69 & 0.09 &  0.06 &  1.25 &  0.06 \\
168740 &        V346 Pav & 365996589 &          1 &    27.78 & 30.75 & 0.00 &   71.12 &   0.14 & 0.03 &  0.02 &  1.84 &  0.03 \\
168947 &        V704 CrA &  89975166 &          1 &    17.24 & 14.59 & 0.00 &  258.05 &   1.64 & 0.06 &  0.04 &  1.00 &  0.05 \\
170625 &        V668 CrA &  91406903 &          1 &    11.36 & 26.56 & 0.00 &  298.55 &   1.41 & 0.12 &  0.06 &  1.22 &  0.06 \\
173794 &        V353 Tel &  324889453 &          0 &    0.31 & 24.87 & 0.00 &  165.15 &   1.53 & 0.04 &  0.06 &  0.99 & 0.11 \\
173844 &             --- & 359678383 &          0 &    15.80 & 15.78 & 0.00 &  178.29 &   0.36 & 0.05 &  0.03 &  2.34 &  0.04 \\
173977 &          HN Dra & 232637376 &          0 &     8.56 &  8.56 & 0.00 &  236.28 &   0.86 & 0.09 &  0.07 &  1.12 &  0.07 \\
176503 &        V544 Lyr &  20816780 &          1 &     8.83 &  8.83 & 0.00 &  271.50 &   1.20 & 0.07 &  0.04 &  0.20 &  0.04 \\
176723 &        V701 CrA & 254079940 &          1 &     7.39 &  8.94 & 0.00 &   65.38 &   0.21 & 0.12 &  0.07 &  1.53 &  0.07 \\
177482 &       sigma Oct & 468184895 &          1 &    10.31 & 10.49 & 0.01 &   90.07 &   0.45 & 0.08 &  0.05 &  0.59 &  0.05 \\
177594 &        V549 Lyr &  42059528 &          1 &     7.85 &  7.85 & 0.01 &  285.15 &   1.44 & 0.04 &  0.03 &  0.77 &  0.03 \\
178905 &             --- & 271523213 &          0 &     8.29 &  8.28 & 0.01 &  359.57 &  29.65 & 0.09 &  0.06 &  0.97 &  0.19 \\
182634 &             --- & 122683109 &          0 &    15.04 & 13.42 & 0.00 &  249.51 &   1.07 & 0.05 &  0.03 &  0.90 &  0.03 \\
184522 &       V2084 Cyg & 138160886 &          1 &    11.36 & 11.37 & 0.01 &  122.89 &   0.22 & 0.09 &  0.05 &  1.81 &  0.05 \\
185139 &          QQ Tel & 143463658 &          1 &    15.31 & 15.54 & 0.01 &  102.07 &   0.31 & 0.06 &  0.04 &  1.15 &  0.04 \\
185332 &       V1745 Cyg & 213679825 &          1 &    18.73 & 25.58 & 0.00 &  140.54 &   0.43 & 0.04 &  0.03 &  1.66 &  0.03 \\
186357 &       V1276 Cyg & 216982075 &          1 &    11.36 & 12.11 & 0.00 &   84.32 &   0.09 & 0.17 &  0.06 &  1.74 &  0.06 \\
186786 &          NZ Pav & 339600902 &          1 &    12.50 & 13.71 & 0.00 &   59.45 &   0.15 & 0.08 &  0.04 &  2.08 &  0.04 \\
187764 &          CN Dra & 258388363 &          1 &    10.00 &  5.05 & 0.01 &  173.06 &   5.93 & 0.12 &  0.09 &  0.02 &  0.11 \\
188136 &          CC Oct & 346793769 &          1 &     8.01 &  8.01 & 0.01 &  193.48 &   3.58 & 0.14 &  0.09 &  1.40 &  0.10 \\
188520 &          CE Oct & 313552150 &          1 &    18.35 & 14.09 & 0.00 &  141.86 &   0.48 & 0.02 &  0.02 &  2.23 &  0.02 \\
191635 &       V2109 Cyg & 294598376 &          1 &     5.37 &  5.38 & 0.05 &  212.02 &   0.68 & 0.09 &  0.07 &  0.78 &  0.07 \\
191804 &          IN Dra & 269697721 &          1 &     7.29 &  7.29 & 0.01 &  217.90 &   1.22 & 0.11 &  0.05 &  1.20 &  0.05 \\
191850 &             --- & 129277739 &          1 &    13.51 & 13.53 & 0.02 &  350.32 &   5.22 & 0.12 &  0.08 &  1.86 &  0.09 \\
192518 &          NU Vul & 452310756 &          1 &     5.32 &  5.32 & 0.01 &   95.98 &   0.55 & 0.04 &  0.03 &  0.25 &  0.03 \\
192640 &       V1644 Cyg &  10988057 &          1 &    37.45 & 37.43 & 0.01 &   40.74 &   0.13 & 0.04 &  0.03 &  1.86 &  0.03 \\
192871 &        V383 Vul & 304501606 &          1 &     5.43 &  7.39 & 0.01 &  199.86 &   0.67 & 0.12 &  0.06 &  0.54 &  0.06 \\
193138 &          IO Dra & 403114672 &          0 &      --- &  8.73 & 0.00 &  185.71 &   3.45 & 0.13 &  0.07 &  1.34 &  0.08 \\
194492 &        V382 Pav & 404239187 &          1 &     7.08 &  7.08 & 0.05 &  385.84 &   1.89 & 0.07 &  0.07 &  0.90 &  0.07 \\
195961 &      $\rho$ Pav & 351535191 &          1 &     8.76 & 10.31 & 0.01 &   62.32 &   0.37 & 0.17 &  0.10 &  0.72 &  0.10 \\
196517 &        V342 Pav & 387241049 &          1 &    29.41 & 32.50 & 0.00 &  167.70 &   0.48 & 0.03 &  0.04 &  2.31 &  0.04 \\
196638 &          BU Mic & 291112607 &          1 &     7.22 &  7.22 & 0.01 &  180.92 &   0.63 & 0.12 &  0.06 &  0.80 &  0.07 \\
197100 &       V2129 Cyg & 343264872 &          1 &     6.46 &  6.46 & 0.02 &  351.17 &   3.40 & 0.08 &  0.06 &  0.50 &  0.06 \\
197157 &         eta Ind & 100708841 &          0 &      --- & 30.47 & 0.00 &   24.20 &   0.06 & 0.04 &  0.04 &  2.54 &  0.04 \\
198830 &          BQ Ind & 354171926 &          1 &    12.20 & 12.20 & 0.08 &  362.89 &   2.34 & 0.06 &  0.05 &  2.00 &  0.06 \\
199434 &        V388 Pav & 372344502 &          1 &     6.32 &  6.32 & 0.01 &  302.80 &   1.48 & 0.07 &  0.06 &  1.28 &  0.06 \\
199757 &          ZZ Mic & 126659093 &          1 &    14.88 & 14.89 & 0.11 &  331.26 &   1.85 & 0.04 &  0.04 &  1.81 &  0.04 \\
199908 &          DQ Cep & 336444537 &          1 &    12.69 & 12.68 & 0.01 &  183.71 &   0.48 & 0.11 &  0.06 &  0.82 &  0.06 \\
200925 &       V1719 Cyg & 290277380 &          1 &     3.74 &  3.74 & 0.09 &  387.29 &   2.32 & 0.10 &  0.08 & -0.05 &  0.09 \\
204615 &      V2455 Cyg  & 266794067 &          0 &    10.62 & 10.62 & 0.13 &     --- &    --- & 0.00 &  0.00 &  1.37 &  0.02 \\
205847 &          CF Ind & 139825582 &          1 &     5.92 &  5.93 & 0.01 &  209.65 &   0.97 & 0.09 &  0.04 &  1.01 &  0.05 \\
206379 &          RS Gru & 139845816 &          1 &     6.80 &  6.80 & 0.15 &  245.74 &   1.18 & 0.05 &  0.05 &  1.26 &  0.05 \\
206553 &          CG Oct & 419666736 &          1 &    15.87 &  8.18 & 0.01 &  103.72 &   0.16 & 0.07 &  0.04 &  1.41 &  0.04 \\
206631 &  V360 Cep       & 322202923 &          1 &    25.00 & 57.53 & 0.00 &     --- &    --- & 0.00 &  0.00 &  2.03 &  0.01 \\
208435 &          BZ Gru & 197686479 &          1 &     6.77 &  8.27 & 0.01 &  125.55 &   0.47 & 0.11 &  0.06 &  0.57 &  0.06 \\
208664 &          BE Ind & 394015973 &          1 &    21.19 & 21.90 & 0.00 &  224.12 &   1.44 & 0.04 &  0.03 &  1.40 &  0.03 \\
208999 &          BX Ind & 265566844 &          1 &     5.63 &  5.63 & 0.03 &  216.69 &   0.77 & 0.10 &  0.07 &  1.11 &  0.07 \\
210111 &             --- & 229059574 &          1 &    27.78 & 30.21 & 0.00 &     --- &    --- & 0.00 &  0.00 &  1.91 &  0.00 \\
211336 & $\epsilon$ Cep  & 330608569 &          1 &    24.27 & 12.73 & 0.00 &     --- &    --- & 0.00 &  0.00 &  2.09 &  0.01 \\
213204 &          UV PsA & 253917376 &          1 &     8.77 &  9.15 & 0.01 &  215.40 &   1.22 & 0.10 &  0.06 &  1.63 &  0.06 \\
213655 &          UW PsA & 209362175 &          1 &    15.38 & 15.71 & 0.00 &  168.40 &   2.22 & 0.11 &  0.05 &  1.26 &  0.06 \\
213669 &      DR Gru     & 219332123 &          0 &    15.02 & 28.24 & 0.01 &     --- &    --- & 0.00 &  0.00 &  2.16 &  0.00 \\
214441 &          CC Gru & 161172103 &          1 &     8.01 &  7.94 & 0.01 &  119.78 &   0.28 & 0.13 &  0.06 &  1.13 &  0.06 \\
217236 &          WX PsA &  89464315 &          1 &     8.01 &  8.01 & 0.01 &   95.33 &   0.63 & 0.08 &  0.07 &  0.52 &  0.08 \\
219586 &        V388 Cep & 427651165 &          1 &     3.68 &  4.30 & 0.01 &  114.92 &   0.49 & 0.07 &  0.05 &  0.18 &  0.05 \\
219891 &             --- & 173771603 &          1 &    10.08 & 10.08 & 0.00 &  137.53 &   0.59 & 0.02 &  0.01 &  0.80 &  0.02 \\
220237 &        V459 Cep & 326201430 &          1 &     5.59 &  5.59 & 0.01 &  210.54 &   0.62 & 0.14 &  0.06 &  0.86 &  0.06 \\
220392 &          DQ Gru & 469933721 &          1 &     4.68 &  4.58 & 0.01 &  125.18 &   0.54 & 0.06 &  0.03 &  0.57 &  0.04 \\
220978 &          BS Scl & 303584611 &          1 &     8.48 & 17.63 & 0.00 &  127.19 &   0.34 & 0.08 &  0.04 &  1.31 &  0.04 \\
221142 &        V377 Cep & 461545977 &          1 &    13.70 & 15.34 & 0.01 &  123.95 &   0.24 & 0.05 &  0.04 &  1.11 &  0.04 \\
221756 &        V340 And & 333375443 &          1 &    22.73 & 22.26 & 0.00 &   76.63 &   0.25 & 0.07 &  0.04 &  1.08 &  0.05 \\
223065 &  SX Phe         & 224285325 &          1 &    18.18 & 18.19 & 0.14 &     --- &    --- & 0.00 &  0.00 &  2.72 &  0.00 \\
223338 &          BS Aqr &   9632550 &          1 &     5.06 &  5.05 & 0.11 &  528.98 &  10.56 & 0.09 &  0.08 &  0.66 &  0.09 \\
223480 &          BF Phe & 144387364 &          1 &    16.03 & 16.02 & 0.00 &  107.13 &   0.22 & 0.05 &  0.03 &  2.22 &  0.03 \\
223661 &        V396 And & 177363174 &          1 &     9.55 &  9.55 & 0.01 &  216.87 &   1.49 & 0.15 &  0.06 &  1.03 &  0.07 \\
224852 &             --- & 355687188 &          0 &     8.19 &  8.19 & 0.07 &  584.00 &   6.92 & 0.10 &  0.07 &  1.02 &  0.08 \\
227695 &       V1821 Cyg &  90350726 &          1 &     9.24 &  8.82 & 0.01 &  607.06 &   7.91 & 0.12 &  0.07 &  0.95 &  0.09 \\
230990 &        V336 Sge & 343216782 &          1 &     5.43 &  6.22 & 0.01 &  474.43 &   5.92 & 0.10 &  0.07 &  0.82 &  0.08 \\
234366 &  V927 Her       & 274510661 &          1 &     7.66 &  7.66 & 0.03 &     --- &    --- & 0.00 &  0.00 &  1.33 &  0.01 \\
254061 &             --- & 166979292 &          0 &    13.33 & 17.22 & 0.08 &  349.58 &   2.67 & 0.05 &  0.05 &  2.36 &  0.07 \\
261331 &        V588 Mon & 220249462 &          1 &     7.14 &  5.14 & 0.00 &  578.77 &   5.33 & 0.05 &  0.03 &  0.91 &  0.06 \\
261446 &        V589 Mon & 220281554 &          1 &     6.27 &  6.49 & 0.01 &  620.57 &   5.24 & 0.14 &  0.08 &  0.90 &  0.08 \\
290764 &       V1247 Ori &  11199521 &          1 &    10.31 & 10.32 & 0.01 &  401.20 &   2.77 & 0.05 &  0.04 &  1.75 &  0.06 \\
290799 &       V1790 Ori &  11361473 &          0 &    23.53 & 22.53 & 0.00 &  420.68 &   3.75 & 0.05 &  0.03 &  2.50 &  0.08 \\
292962 &             --- &  33149129 &          0 &     9.19 &  9.19 & 0.07 &  486.18 &   3.75 & 0.06 &  0.05 &  1.41 &  0.06 \\
316092 &        V974 Oph & 200624064 &          1 &     5.23 &  5.23 & 0.07 & 1088.42 &  73.67 & 0.28 &  0.17 &  1.04 &  0.27 \\
339660 &        V381 Vul & 245151597 &          1 &    17.86 & 10.43 & 0.00 &  814.38 &  28.26 & 0.06 &  0.06 &  0.56 &  0.10 \\
341555 &        V979 Her & 135362719 &          1 &     8.65 &  8.65 & 0.01 &  230.41 &   0.81 & 0.08 &  0.06 &  1.56 &  0.06 \\
   --- &          AB Cas & 354922610 &          1 &    17.15 & 17.16 & 0.01 &  329.74 &   1.47 & 0.11 &  0.10 &  2.59 &  0.11 \\
   --- &          AE UMa & 357132618 &          1 &    11.63 & 11.62 & 0.14 &  735.57 &  16.08 & 0.09 &  0.07 &  1.93 &  0.12 \\
   --- &          AI Hya & 455178154 &          1 &     7.25 &  6.24 & 0.00 &  611.24 &   6.20 & 0.12 &  0.07 &  0.31 &  0.07 \\
   --- &          AN Lyn &  56882581 &          1 &    10.17 & 10.17 & 0.05 &  666.20 &  10.31 & 0.09 &  0.07 &  1.44 &  0.09 \\
   --- &          BO Lyn &  99091734 &          1 &    10.71 & 10.71 & 0.06 & 1343.42 &  29.97 & 0.06 &  0.06 &  0.79 &  0.15 \\
   --- &          CC And & 191466237 &          1 &     8.01 &  8.00 & 0.04 &  379.57 &   2.26 & 0.08 &  0.07 &  1.37 &  0.07 \\
   --- &          DE Lac & 119486942 &          1 &     3.94 &  4.00 & 0.03 &  765.00 &  10.25 & 0.17 &  0.12 &  0.75 &  0.12 \\
   --- &          ES Cir & 298112357 &          0 &     9.84 &  9.85 & 0.17 &  595.38 &   5.16 & 0.12 &  0.10 &  1.96 &  0.12 \\
   --- &          GP And & 436546358 &          1 &    12.71 & 12.71 & 0.16 &  545.24 &   7.66 & 0.09 &  0.08 &  2.03 &  0.13 \\
   --- &          GP Cnc & 458658640 &          0 &     8.84 & 11.77 & 0.01 &     --- &    --- & 0.00 &  0.00 &  2.04 &  0.20 \\
   --- &          GW Dra & 329153513 &          1 &     7.92 &  7.92 & 0.02 &  358.69 &   2.04 & 0.10 &  0.07 &  1.40 &  0.08 \\
   --- &          GW UMa & 150276417 &          0 &     4.92 &  4.92 & 0.12 &  676.48 &  21.37 & 0.20 &  0.13 &  0.34 &  0.14 \\
   --- &          OS Gem & 171300396 &          1 &    15.48 & 14.50 & 0.00 &  453.16 &   2.79 & 0.04 &  0.03 &  1.06 &  0.04 \\
   --- &          UV Tri &  61275530 &          1 &    10.00 &  9.33 & 0.01 &  710.73 &  15.91 & 0.08 &  0.06 &  1.81 &  0.12 \\
   --- &       V1162 Ori &  34512862 &          1 &    12.71 & 12.71 & 0.04 &  444.01 &   4.65 & 0.08 &  0.07 &  1.67 &  0.08 \\
   --- &       V1199 Ori &  43846109 &          0 &      --- & 13.03 & 0.00 &  608.61 &   6.48 & 0.10 &  0.05 &  2.69 &  0.21 \\
   --- &       V1232 Cen & 181087723 &          0 &      --- &  7.72 & 0.11 & 1450.15 &  25.54 & 0.11 &  0.09 &  1.33 &  0.10 \\
   --- &        V402 Cep & 468294062 &          1 &     8.14 &  8.13 & 0.01 &  630.06 &   4.41 & 0.15 &  0.09 &  1.40 &  0.10 \\
   --- &        V420 Car & 364398340 &          0 &      --- & 16.03 & 0.00 &  423.76 &   8.31 & 0.07 &  0.04 &  2.18 &  0.05 \\
   --- &        V421 Car & 364398097 &          0 &      --- & 39.12 & 0.00 &  411.41 &   1.82 & 0.05 &  0.03 &  2.75 &  0.03 \\
   --- &        V459 Per & 259254245 &          1 &    27.03 & 22.48 & 0.00 &  174.82 &   0.44 & 0.11 &  0.05 &  2.86 &  0.05 \\
   --- &        V593 Lyr & 289325437 &          0 &    11.36 &  9.79 & 0.18 & 1459.59 &  25.88 & 0.09 &  0.07 &  0.77 &  0.19 \\
   --- &        V673 Hya & 168384036 &          0 &     9.25 &  9.25 & 0.11 &  576.54 &   7.62 & 0.09 &  0.08 &  1.73 &  0.10 \\
   --- &        V798 Cyg &  69546708 &          1 &     5.13 &  5.13 & 0.14 & 2450.28 &  51.23 & 0.09 &  0.07 &  0.96 &  0.08 \\
   --- &        V830 Her & 308407544 &          1 &     5.56 &  5.65 & 0.03 &  466.45 &   2.58 & 0.12 &  0.09 &  0.82 &  0.09 \\
   --- &        V871 Cas & 252046615 &          1 &     7.71 &  7.71 & 0.03 &     --- &    --- & 0.00 &  0.00 &  1.32 &  0.02 \\
   --- &          VX Hya & 289711518 &          1 &     4.48 &  4.48 & 0.09 &  932.72 &  12.68 & 0.07 &  0.07 &  0.68 &  0.11 \\
   --- &          XX Cyg & 233310793 &          1 &     7.41 &  7.42 & 0.23 & 1162.39 &  20.23 & 0.11 &  0.09 &  1.43 &  0.19 \\
   --- &          XX Pyx &  37366830 &          1 &    38.17 & 38.11 & 0.01 &  703.30 &  13.90 & 0.06 &  0.04 &  2.21 &  0.06 \\
   --- &  Y Cam          & 441598163 &          1 &    15.04 & 15.05 & 0.00 &  799.37 &   6.46 & 0.09 &  0.10 &  1.00 &  0.10 \\
   --- &          YZ Boo & 233465540 &          1 &     9.61 &  9.61 & 0.11 &  585.02 &   4.71 & 0.06 &  0.05 &  1.62 &  0.07 \\
   --- &             --- &  13972612 &          0 &     9.60 & 10.61 & 0.00 & 1778.09 &  26.62 & 0.22 &  0.10 & -0.49 &  0.13 \\
   --- &             --- &  46937596 &          0 &     7.89 &  7.89 & 0.07 &  811.80 &  10.99 & 0.09 &  0.08 &  1.54 &  0.15 \\
   --- &             --- &  60405689 &          0 &    14.99 & 14.99 & 0.06 &  567.84 &   5.83 & 0.08 &  0.06 &  2.48 &  0.11 \\
   --- &             --- &  65138566 &          0 &    18.35 & 18.33 & 0.03 &  538.03 &   5.08 & 0.06 &  0.05 &  2.48 &  0.10 \\
   --- &             --- &  78850814 &          0 &     9.12 &  9.12 & 0.07 & 1099.34 &  10.85 & 0.15 &  0.11 &  1.25 &  0.17 \\
   --- &             --- &  90185615 &          0 &     7.20 &  7.20 & 0.06 &  578.33 &   5.45 & 0.13 &  0.12 &  1.63 &  0.14 \\
   --- &             --- &  90322352 &          0 &     9.33 &  9.33 & 0.10 & 1074.32 &  29.73 & 0.08 &  0.08 &  1.29 &  0.17 \\
   --- &             --- & 120857354 &          0 &    20.49 & 20.49 & 0.04 &     --- &    --- & 0.00 &  0.00 &  2.46 &  0.01 \\
   --- &             --- & 130474019 &          0 &      --- & 12.28 & 0.12 &  975.66 &  10.18 & 0.08 &  0.07 &  2.20 &  0.22 \\
   --- &             --- & 142945544 &          0 &    16.47 & 18.48 & 0.07 &  463.26 &   9.86 & 0.05 &  0.05 &  2.40 &  0.09 \\
   --- &             --- & 148357344 &          0 &      --- & 11.10 & 0.14 & 1229.90 &  21.70 & 0.15 &  0.11 &  1.51 &  0.28 \\
   --- &             --- & 155128092 &          0 &      --- & 14.68 & 0.10 &  775.74 &  12.54 & 0.06 &  0.05 &  1.90 &  0.11 \\
   --- &             --- & 178616716 &          0 &      --- &  9.76 & 0.09 & 1224.71 &  23.01 & 0.11 &  0.10 &  1.43 &  0.10 \\
   --- &             --- & 183532876 &          0 &     5.98 &  5.98 & 0.16 &     --- &    --- & 0.00 &  0.00 &  0.95 &  0.05 \\
   --- &             --- & 261089835 &          0 &     7.15 & 11.16 & 0.08 & 1226.34 &  28.75 & 0.14 &  0.11 &  1.32 &  0.21 \\
   --- &             --- & 262652067 &          0 &     5.63 &  5.63 & 0.06 &  726.83 &   5.89 & 0.15 &  0.11 &  1.37 &  0.13 \\
   --- &             --- & 264482621 &          0 &    18.80 & 18.78 & 0.04 &  392.61 &   2.46 & 0.08 &  0.09 &  2.71 &  0.13 \\
   --- &             --- & 278003738 &          0 &     8.55 &  8.55 & 0.02 &  560.78 &   4.72 & 0.09 &  0.07 &  1.89 &  0.11 \\
   --- &             --- & 308396022 &          0 &    13.21 & 13.20 & 0.07 &  600.99 &  10.85 & 0.10 &  0.09 &  2.01 &  0.13 \\
   --- &             --- & 308456254 &          0 &     6.28 &  6.28 & 0.04 &  706.94 &  11.11 & 0.11 &  0.09 &  1.57 &  0.12 \\
   --- &             --- & 337437469 &          0 &     7.62 &  7.62 & 0.03 &  725.73 &   5.51 & 0.11 &  0.11 &  1.60 &  0.11 \\
   --- &             --- & 339675323 &          0 &     7.45 &  7.45 & 0.06 &  936.41 &  18.04 & 0.16 &  0.12 &  1.58 &  0.19 \\
   --- &             --- & 349972148 &          0 &     5.93 &  5.93 & 0.04 &     --- &    --- & 0.00 &  0.00 &  1.37 &  0.02 \\
   --- &             --- & 358502706 &          0 &    11.67 & 11.67 & 0.08 &  984.78 &  11.30 & 0.08 &  0.06 &  1.73 &  0.07 \\
   --- &             --- & 360736543 &          0 &     6.71 &  6.71 & 0.05 & 1393.52 &  29.54 & 0.08 &  0.07 &  0.82 &  0.14 \\
   --- &             --- & 362384415 &          0 &      --- &  7.22 & 0.13 & 1161.49 &  30.53 & 0.11 &  0.09 &  1.68 &  0.22 \\
   --- &             --- & 374753270 &          0 &     7.25 &  7.25 & 0.15 &  733.07 &   4.85 & 0.12 &  0.11 &  1.61 &  0.11 \\
   --- &             --- & 391474560 &          0 &      --- & 10.54 & 0.09 & 1122.74 &  11.83 & 0.10 &  0.08 &  1.70 &  0.24 \\
   --- &             --- & 393420032 &          0 &     6.01 &  6.01 & 0.06 & 1204.71 &  25.86 & 0.15 &  0.10 &  0.69 &  0.13 \\
   --- &             --- & 410038602 &          0 &      --- & 13.07 & 0.06 & 1163.57 &  23.90 & 0.06 &  0.05 &  1.92 &  0.21 \\
   --- &             --- & 431589510 &          0 &      --- &  7.73 & 0.08 & 1356.52 & 138.29 & 0.10 &  0.09 &  1.60 &  0.31 \\
   --- &             --- & 453717216 &          0 &     8.35 &  8.35 & 0.02 &     --- &    --- & 0.00 &  0.00 &  2.31 &  0.01 \\

%% file: paper_final.bbl
\begin{thebibliography}{}
\makeatletter
\relax
\def\mn@urlcharsother{\let\do\@makeother \do\$\do\&\do\#\do\^\do\_\do\%\do\~}
\def\mn@doi{\begingroup\mn@urlcharsother \@ifnextchar [ {\mn@doi@}
  {\mn@doi@[]}}
\def\mn@doi@[#1]#2{\def\@tempa{#1}\ifx\@tempa\@empty \href
  {http://dx.doi.org/#2} {doi:#2}\else \href {http://dx.doi.org/#2} {#1}\fi
  \endgroup}
\def\mn@eprint#1#2{\mn@eprint@#1:#2::\@nil}
\def\mn@eprint@arXiv#1{\href {http://arxiv.org/abs/#1} {{\tt arXiv:#1}}}
\def\mn@eprint@dblp#1{\href {http://dblp.uni-trier.de/rec/bibtex/#1.xml}
  {dblp:#1}}
\def\mn@eprint@#1:#2:#3:#4\@nil{\def\@tempa {#1}\def\@tempb {#2}\def\@tempc
  {#3}\ifx \@tempc \@empty \let \@tempc \@tempb \let \@tempb \@tempa \fi \ifx
  \@tempb \@empty \def\@tempb {arXiv}\fi \@ifundefined
  {mn@eprint@\@tempb}{\@tempb:\@tempc}{\expandafter \expandafter \csname
  mn@eprint@\@tempb\endcsname \expandafter{\@tempc}}}

\bibitem[\protect\citeauthoryear{{Aerts}}{{Aerts}}{2021}]{Aerts2021}
{Aerts} C.,  2021, \mn@doi [Reviews of Modern Physics]
  {10.1103/RevModPhys.93.015001}, \href
  {https://ui.adsabs.harvard.edu/abs/2021RvMP...93a5001A} {93, 015001}

\bibitem[\protect\citeauthoryear{{Aerts}, {Handler}, {Arentoft},
  {Vandenbussche}, {Medupe}  \& {Sterken}}{{Aerts} et~al.}{2002}]{Aerts2002}
{Aerts} C.,  {Handler} G.,  {Arentoft} T.,  {Vandenbussche} B.,  {Medupe} R.,
  {Sterken} C.,  2002, \mn@doi [\mnras] {10.1046/j.1365-8711.2002.05627.x},
  \href {https://ui.adsabs.harvard.edu/abs/2002MNRAS.333L..35A} {333, L35}

\bibitem[\protect\citeauthoryear{{Antonello} \& {Pasinetti
  Fracassini}}{{Antonello} \& {Pasinetti
  Fracassini}}{1998}]{Antonello+Pasinetti-Fracassini1998}
{Antonello} E.,  {Pasinetti Fracassini} L.~E.,  1998, \aap, \href
  {https://ui.adsabs.harvard.edu/abs/1998A&A...331..995A} {331, 995}

\bibitem[\protect\citeauthoryear{{Bedding} et~al.,}{{Bedding}
  et~al.}{2020}]{Bedding2020}
{Bedding} T.~R.,  et~al., 2020, \mn@doi [\nat] {10.1038/s41586-020-2226-8},
  \href {https://ui.adsabs.harvard.edu/abs/2020Natur.581..147B} {581, 147}

\bibitem[\protect\citeauthoryear{{Berger}, {Huber}, {van Saders}, {Gaidos},
  {Tayar}  \& {Kraus}}{{Berger} et~al.}{2020}]{Berger2020}
{Berger} T.~A.,  {Huber} D.,  {van Saders} J.~L.,  {Gaidos} E.,  {Tayar} J.,
  {Kraus} A.~L.,  2020, \mn@doi [\aj] {10.3847/1538-3881/159/6/280}, \href
  {https://ui.adsabs.harvard.edu/abs/2020AJ....159..280B} {159, 280}

\bibitem[\protect\citeauthoryear{{Bowman}, {Kurtz}, {Breger}, {Murphy}  \&
  {Holdsworth}}{{Bowman} et~al.}{2016}]{Bowman2016}
{Bowman} D.~M.,  {Kurtz} D.~W.,  {Breger} M.,  {Murphy} S.~J.,   {Holdsworth}
  D.~L.,  2016, \mn@doi [\mnras] {10.1093/mnras/stw1153}, \href
  {http://adsabs.harvard.edu/abs/2016MNRAS.460.1970B} {460, 1970}

\bibitem[\protect\citeauthoryear{{Breger}}{{Breger}}{2000}]{Breger2000}
{Breger} M.,  2000, in ASP Conf. Series, Vol.~210. p.~3

\bibitem[\protect\citeauthoryear{{Casas}, {Moya}, {Su{\'a}rez},
  {Mart{\'\i}n-Ru{\'\i}z}, {Amado}, {Rodr{\'\i}guez-L{\'o}pez}  \&
  {Garrido}}{{Casas} et~al.}{2009}]{Casas2009}
{Casas} R.,  {Moya} A.,  {Su{\'a}rez} J.~C.,  {Mart{\'\i}n-Ru{\'\i}z} S.,
  {Amado} P.~J.,  {Rodr{\'\i}guez-L{\'o}pez} C.,   {Garrido} R.,  2009, \mn@doi
  [\apj] {10.1088/0004-637X/697/1/522}, \href
  {https://ui.adsabs.harvard.edu/abs/2009ApJ...697..522C} {697, 522}

\bibitem[\protect\citeauthoryear{{Castelli} \& {Kurucz}}{{Castelli} \&
  {Kurucz}}{2003}]{castelli&kurucz2003}
{Castelli} F.,  {Kurucz} R.~L.,  2003, in {Piskunov} N.,  {Weiss} W.~W.,
  {Gray} D.~F.,  eds,  IAU Symposium Vol. 210, Modelling of Stellar
  Atmospheres. p.~A20 (\mn@eprint {arXiv} {astro-ph/0405087})

\bibitem[\protect\citeauthoryear{{Catelan} \& {Smith}}{{Catelan} \&
  {Smith}}{2015}]{Catelan+Smith2015}
{Catelan} M.,  {Smith} H.~A.,  2015, {Pulsating Stars}.
Wiley-VCH, Weinheim, Germany

\bibitem[\protect\citeauthoryear{{Chang}, {Protopapas}, {Kim}  \&
  {Byun}}{{Chang} et~al.}{2013}]{Chang2013}
{Chang} S.~W.,  {Protopapas} P.,  {Kim} D.~W.,   {Byun} Y.~I.,  2013, \mn@doi
  [\aj] {10.1088/0004-6256/145/5/132}, \href
  {https://ui.adsabs.harvard.edu/abs/2013AJ....145..132C} {145, 132}

\bibitem[\protect\citeauthoryear{ESA}{ESA}{1997}]{ESA1997}
ESA 1997, {The HIPPARCOS and TYCHO catalogues. Astrometric and photometric star
  catalogues derived from the ESA HIPPARCOS Space Astrometry Mission}.
 ESA Special Publication Vol. 1200, ESA Publications Division, Noordwijk,
  Netherlands

\bibitem[\protect\citeauthoryear{{Fitch}}{{Fitch}}{1981}]{Fitch1981}
{Fitch} W.~S.,  1981, \mn@doi [\apj] {10.1086/159278}, \href
  {https://ui.adsabs.harvard.edu/abs/1981ApJ...249..218F} {249, 218}

\bibitem[\protect\citeauthoryear{{Gaia Collaboration} et~al.,}{{Gaia
  Collaboration} et~al.}{2022}]{Gaia-De-Ridder++2022}
{Gaia Collaboration} et~al., 2022, arXiv e-prints, \href
  {https://ui.adsabs.harvard.edu/abs/2022arXiv220606075G} {p. arXiv:2206.06075}

\bibitem[\protect\citeauthoryear{{Garofalo}, {Delgado}, {Sarro}, {Clementini},
  {Muraveva}, {Marconi}  \& {Ripepi}}{{Garofalo} et~al.}{2022}]{Garofalo++2022}
{Garofalo} A.,  {Delgado} H.~E.,  {Sarro} L.~M.,  {Clementini} G.,  {Muraveva}
  T.,  {Marconi} M.,   {Ripepi} V.,  2022, arXiv e-prints, \href
  {https://ui.adsabs.harvard.edu/abs/2022arXiv220307435G} {p. arXiv:2203.07435}

\bibitem[\protect\citeauthoryear{{Guzik}}{{Guzik}}{2021}]{Guzik2021}
{Guzik} J.~A.,  2021, \mn@doi [Frontiers in Astronomy and Space Sciences]
  {10.3389/fspas.2021.653558}, \href
  {https://ui.adsabs.harvard.edu/abs/2021FrASS...8...55G} {8, 55}

\bibitem[\protect\citeauthoryear{{Handler} et~al.,}{{Handler}
  et~al.}{2000}]{Handler++2000}
{Handler} G.,  et~al., 2000, \mn@doi [\mnras]
  {10.1046/j.1365-8711.2000.03817.x}, \href
  {https://ui.adsabs.harvard.edu/abs/2000MNRAS.318..511H} {318, 511}

\bibitem[\protect\citeauthoryear{{Huber} et~al.,}{{Huber}
  et~al.}{2010}]{Huber2010}
{Huber} D.,  et~al., 2010, \mn@doi [\apj] {10.1088/0004-637X/723/2/1607}, \href
  {http://adsabs.harvard.edu/abs/2010ApJ...723.1607H} {723, 1607}

\bibitem[\protect\citeauthoryear{{Jayasinghe} et~al.,}{{Jayasinghe}
  et~al.}{2020}]{Jayasinghe2020}
{Jayasinghe} T.,  et~al., 2020, \mn@doi [\mnras] {10.1093/mnras/staa499}, \href
  {https://ui.adsabs.harvard.edu/abs/2020MNRAS.493.4186J} {493, 4186}

\bibitem[\protect\citeauthoryear{{Kurtz}}{{Kurtz}}{2022}]{Kurtz2022}
{Kurtz} D.,  2022, arXiv e-prints, \href
  {https://ui.adsabs.harvard.edu/abs/2022arXiv220111629K} {p. arXiv:2201.11629}

\bibitem[\protect\citeauthoryear{{Leavitt} \& {Pickering}}{{Leavitt} \&
  {Pickering}}{1912}]{Leavitt+Pickering1912}
{Leavitt} H.~S.,  {Pickering} E.~C.,  1912, Harvard College Observatory
  Circular, \href {http://adsabs.harvard.edu/abs/1912HarCi.173....1L} {173, 1}

\bibitem[\protect\citeauthoryear{{Li}, {Qian}, {Zhang}, {Zhu}  \& {Liao}}{{Li}
  et~al.}{2018}]{Li2018}
{Li} L.-J.,  {Qian} S.-B.,  {Zhang} J.,  {Zhu} L.-Y.,   {Liao} W.-P.,  2018,
  \mn@doi [Research in Astronomy and Astrophysics] {10.1088/1674-4527/18/1/11},
  \href {https://ui.adsabs.harvard.edu/abs/2018RAA....18...11L} {18, 011}

\bibitem[\protect\citeauthoryear{{Li}, {Van Reeth}, {Bedding}, {Murphy},
  {Antoci}, {Ouazzani}  \& {Barbara}}{{Li} et~al.}{2020}]{Li++2020}
{Li} G.,  {Van Reeth} T.,  {Bedding} T.~R.,  {Murphy} S.~J.,  {Antoci} V.,
  {Ouazzani} R.-M.,   {Barbara} N.~H.,  2020, \mn@doi [\mnras]
  {10.1093/mnras/stz2906}, \href
  {https://ui.adsabs.harvard.edu/abs/2020MNRAS.491.3586L} {491, 3586}

\bibitem[\protect\citeauthoryear{{Liakos} \& {Niarchos}}{{Liakos} \&
  {Niarchos}}{2017}]{Liakos+Niarchos2017}
{Liakos} A.,  {Niarchos} P.,  2017, \mn@doi [\mnras] {10.1093/mnras/stw2756},
  \href {http://adsabs.harvard.edu/abs/2017MNRAS.465.1181L} {465, 1181}

\bibitem[\protect\citeauthoryear{{Lightkurve Collaboration}
  et~al.,}{{Lightkurve Collaboration}
  et~al.}{2018}]{lightkurvecollaboration2018}
{Lightkurve Collaboration} et~al., 2018, {Lightkurve: Kepler and TESS time
  series analysis in Python}, Astrophysics Source Code Library (\mn@eprint
  {ascl} {1812.013})

\bibitem[\protect\citeauthoryear{{Lovekin} \& {Guzik}}{{Lovekin} \&
  {Guzik}}{2017}]{Lovekin+Guzik2017}
{Lovekin} C.~C.,  {Guzik} J.~A.,  2017, \mn@doi [\apj]
  {10.3847/1538-4357/aa8e01}, \href
  {https://ui.adsabs.harvard.edu/abs/2017ApJ...849...38L} {849, 38}

\bibitem[\protect\citeauthoryear{{McNamara}}{{McNamara}}{1997}]{McNamara1997}
{McNamara} D.,  1997, \mn@doi [\pasp] {10.1086/133999}, \href
  {http://adsabs.harvard.edu/abs/1997PASP..109.1221M} {109, 1221}

\bibitem[\protect\citeauthoryear{{McNamara}}{{McNamara}}{2011}]{McNamara2011}
{McNamara} D.~H.,  2011, \mn@doi [\aj] {10.1088/0004-6256/142/4/110}, \href
  {http://adsabs.harvard.edu/abs/2011AJ....142..110M} {142, 110}

\bibitem[\protect\citeauthoryear{{Mkrtichian} et~al.,}{{Mkrtichian}
  et~al.}{2007}]{Mkrtichian2007}
{Mkrtichian} D.~E.,  et~al., 2007, \mn@doi [\aj] {10.1086/521434}, \href
  {https://ui.adsabs.harvard.edu/abs/2007AJ....134.1713M} {134, 1713}

\bibitem[\protect\citeauthoryear{{Moln{\'a}r} et~al.,}{{Moln{\'a}r}
  et~al.}{2022}]{Molnar++2022}
{Moln{\'a}r} L.,  et~al., 2022, \mn@doi [\apjs] {10.3847/1538-4365/ac2ee2},
  \href {https://ui.adsabs.harvard.edu/abs/2022ApJS..258....8M} {258, 8}

\bibitem[\protect\citeauthoryear{{Murphy}}{{Murphy}}{2012}]{murphy2012}
{Murphy} S.~J.,  2012, \mn@doi [\mnras] {10.1111/j.1365-2966.2012.20644.x},
  \href {http://adsabs.harvard.edu/abs/2012MNRAS.422..665M} {422, 665}

\bibitem[\protect\citeauthoryear{{Murphy}, {Hey}, {Van Reeth}  \&
  {Bedding}}{{Murphy} et~al.}{2019}]{Murphy2019}
{Murphy} S.~J.,  {Hey} D.,  {Van Reeth} T.,   {Bedding} T.~R.,  2019, \mn@doi
  [\mnras] {10.1093/mnras/stz590}, \href
  {https://ui.adsabs.harvard.edu/abs/2019MNRAS.485.2380M} {485, 2380}

\bibitem[\protect\citeauthoryear{{Murphy}, {Paunzen}, {Bedding}, {Walczak}  \&
  {Huber}}{{Murphy} et~al.}{2020}]{Murphy2020}
{Murphy} S.~J.,  {Paunzen} E.,  {Bedding} T.~R.,  {Walczak} P.,   {Huber} D.,
  2020, \mn@doi [\mnras] {10.1093/mnras/staa1271}, \href
  {https://ui.adsabs.harvard.edu/abs/2020MNRAS.495.1888M} {495, 1888}

\bibitem[\protect\citeauthoryear{{Murphy}, {Bedding}, {White}, {Li}, {Hey},
  {Reese}  \& {Joyce}}{{Murphy} et~al.}{2022}]{murphy2022}
{Murphy} S.~J.,  {Bedding} T.~R.,  {White} T.~R.,  {Li} Y.,  {Hey} D.,  {Reese}
  D.,   {Joyce} M.,  2022, \mn@doi [\mnras] {10.1093/mnras/stac240}, \href
  {https://ui.adsabs.harvard.edu/abs/2022MNRAS.511.5718M} {511, 5718}

\bibitem[\protect\citeauthoryear{{North}, {Jaschek}  \& {Egret}}{{North}
  et~al.}{1997}]{North1997}
{North} P.,  {Jaschek} C.,   {Egret} D.,  1997, in {Bonnet} R.~M.,  et~al.,
  eds,  ESA Special Publication Vol. 402, Hipparcos -- Venice '97. p.~367

\bibitem[\protect\citeauthoryear{{Ochsenbein}, {Bauer}  \&
  {Marcout}}{{Ochsenbein} et~al.}{2000}]{vizier}
{Ochsenbein} F.,  {Bauer} P.,   {Marcout} J.,  2000, \mn@doi [\aaps]
  {10.1051/aas:2000169}, \href
  {http://adsabs.harvard.edu/abs/2000A%26AS..143...23O} {143, 23}

\bibitem[\protect\citeauthoryear{{Petersen} \&
  {Christensen-Dalsgaard}}{{Petersen} \&
  {Christensen-Dalsgaard}}{1996}]{Petersen+CD1996}
{Petersen} J.~O.,  {Christensen-Dalsgaard} J.,  1996, \aap, \href
  {https://ui.adsabs.harvard.edu/abs/1996A&A...312..463P} {312, 463}

\bibitem[\protect\citeauthoryear{{Pigulski} \& {Pojma{\'n}ski}}{{Pigulski} \&
  {Pojma{\'n}ski}}{2008}]{Pigulski2008}
{Pigulski} A.,  {Pojma{\'n}ski} G.,  2008, \mn@doi [\aap]
  {10.1051/0004-6361:20078580}, \href
  {http://adsabs.harvard.edu/abs/2008A%26A...477..907P} {477, 907}

\bibitem[\protect\citeauthoryear{{Poro} et~al.,}{{Poro}
  et~al.}{2021}]{Poro2021}
{Poro} A.,  et~al., 2021, \mn@doi [\pasp] {10.1088/1538-3873/ac12dc}, \href
  {https://ui.adsabs.harvard.edu/abs/2021PASP..133h4201P} {133, 084201}

\bibitem[\protect\citeauthoryear{{Rodr{\'{\i}}guez}, {L{\'o}pez-Gonz{\'a}lez}
  \& {L{\'o}pez de Coca}}{{Rodr{\'{\i}}guez} et~al.}{2000}]{Rodriguez2000}
{Rodr{\'{\i}}guez} E.,  {L{\'o}pez-Gonz{\'a}lez} M.~J.,   {L{\'o}pez de Coca}
  P.,  2000, \aaps, \href {http://adsabs.harvard.edu/abs/2000A%26AS..144..469R}
  {144, 469}

\bibitem[\protect\citeauthoryear{{Sneden} et~al.,}{{Sneden}
  et~al.}{2018}]{Sneden2018}
{Sneden} C.,  et~al., 2018, \mn@doi [\aj] {10.3847/1538-3881/aa9f16}, \href
  {http://adsabs.harvard.edu/abs/2018AJ....155...45S} {155, 45}

\bibitem[\protect\citeauthoryear{{Speagle}}{{Speagle}}{2020}]{speagle2020}
{Speagle} J.~S.,  2020, \mn@doi [\mnras] {10.1093/mnras/staa278}, \href
  {https://ui.adsabs.harvard.edu/abs/2020MNRAS.493.3132S} {493, 3132}

\bibitem[\protect\citeauthoryear{{Steindl}, {Zwintz}  \& {Bowman}}{{Steindl}
  et~al.}{2021}]{steindletal2021}
{Steindl} T.,  {Zwintz} K.,   {Bowman} D.~M.,  2021, \mn@doi [\aap]
  {10.1051/0004-6361/202039093}, \href
  {https://ui.adsabs.harvard.edu/abs/2021A&A...645A.119S} {645, A119}

\bibitem[\protect\citeauthoryear{Terrell \& Scott}{Terrell \&
  Scott}{1992}]{Terrell+Scott1992}
Terrell G.~R.,  Scott D.~W.,  1992, \mn@doi [The Annals of Statistics]
  {10.1214/aos/1176348768}, 20, 1236

\bibitem[\protect\citeauthoryear{{Ziaali}, {Bedding}, {Murphy}, {Van Reeth}  \&
  {Hey}}{{Ziaali} et~al.}{2019}]{Ziaali2019}
{Ziaali} E.,  {Bedding} T.~R.,  {Murphy} S.~J.,  {Van Reeth} T.,   {Hey} D.~R.,
   2019, \mn@doi [\mnras] {10.1093/mnras/stz1110}, \href
  {https://ui.adsabs.harvard.edu/abs/2019MNRAS.486.4348Z} {486, 4348}

\makeatother
\end{thebibliography}
